\documentclass{emulateapj}
\usepackage{rotating}
\usepackage{epsf}
\usepackage{psfig}
\usepackage{float}
\usepackage [english]{babel}
\usepackage [autostyle, english = american]{csquotes}
\usepackage{chngcntr}
\usepackage{multirow}

\MakeOuterQuote{"}

\begin{document}

\title{Powerful H$_2$ Line Cooling in Stephan's Quintet II. Group-wide Gas and Shock Modeling of the Warm H$_2$ and a comparison with [\ion{C}{2}]157.7$\mu$m emission and kinematics} 

\author{P. N. Appleton\altaffilmark{1}, P. Guillard\altaffilmark{2},  A. Togi\altaffilmark{3},  K. Alatalo\altaffilmark{4}, F. Boulanger\altaffilmark{5}, M. Cluver\altaffilmark{6}, G. Pineau des For{\^e}ts\altaffilmark{5,7}, U. Lisenfeld\altaffilmark{8}, P. Ogle\altaffilmark{9} \& C. K. Xu\altaffilmark{1,9}.}

\altaffiltext{1}{NASA {\it Herschel} Science Center, IPAC, Caltech, 770S Wilson Av., Pasadena, CA 91125. apple@ipac.caltech.edu}
\altaffiltext{2}{Sorbonne Universit�s,  UPMC Univ Paris 6 et CNRS, UMR 7095, Institut d'Astrophysique de Paris, 98 bis Bd Arago, 75014 Paris, France}
\altaffiltext{3}{Department of Physics and Astronomy, The University of Toledo, 2825 West Bancroft Street, Toledo, OH 43606}
\altaffiltext{4}{Observatories of the Carnegie Institution for Science, 813 Santa Barbara Street, Pasadena, CA 91101, Hubble Fellow}
\altaffiltext{5}{Institut d'Astrophysique Spatiale, 91405 Orsay, Universite Paris Sud et CNRS, France }
\altaffiltext{6}{Department of Physics and Astronomy,University of Western Cape, Robert Sobukwe Rd., Bellville, 7535, South Africa}
\altaffiltext{7}{DAEC, Observatoire de Paris, F-92195 Meudon Principal Cedex, France.}
\altaffiltext{8}{Departamento de F\'isica Teorica y del Cosmos, Universidad de Granada, Spain and Instituto Carlos I de F\'isica Teorica y
Computacional, Facultad de Ciencias, 18071 Granada, Spain}
\altaffiltext{9}{NASA Extragalactic Database, IPAC, Caltech, 1200 E. California Blvd, Caltech, Pasadena, CA 91125 }

\begin{abstract}
We map for the first time the two-dimensional H$_2$ excitation of warm intergalactic gas in Stephan'��s Quintet on group-wide (50 x 35 kpc$^2$) scales to quantify the temperature, mass and warm-H$_2$ mass fraction as a function of position using {\it Spitzer}.  Molecular gas temperatures are seen to rise (to T $>$ 700K) and the slope of the power--law density-temperature relation flattens along the main ridge of the filament, defining the region of maximum heating. We also performed MHD modeling of the excitation properties of the warm gas, to map the velocity structure and energy deposition rate of slow and fast molecular shocks. Slow magnetic shocks were required to explain the power radiated from the lowest--lying rotational states of H$_2$, and strongly support the idea that energy cascades down to small scales and low velocities from the fast collision of NGC 7318b with group-wide gas.    The highest levels of heating of the warm H$_2$ is strongly correlated with the large-scale stirring of the medium as measured by [\ion{C}{2}] spectroscopy with {\it Herschel}.  H$_2$ is also seen associated with a separate bridge which extends towards the Seyfert nucleus in NGC 7319, both from {\it Spitzer} and {\it CARMA} CO observations. This opens up the possibility that both galaxy collisions and AGN outflows can turbulently heat gas on large-scales in compact groups. The observations provide a laboratory for studying the effects of turbulent energy dissipation on group-wide scales that may provide clues about the heating and cooling of gas at high-z in early galaxy and protogalaxy formation.
\end{abstract}


\section{Introduction}

Stephan's Quintet (hereafter the Quintet) is one of the most studied of nearby compact groups of galaxies. Observations have been made across a wide range of the electromagnetic spectrum, from X-rays \citep{bah84,sul95,tri03,osu09}, UV \citep{xu05}, visible light \citep{mol98,gal01,fed11}, IR \citep{sul01,nat10,clu10,gui10,suz11,bit14} to radio waves \citep{all72,van81,xu03,nik13}. 
These papers have revealed that much of the gas lies either in tidal tails outside the galaxies, or in a large intergalactic filament.  The filament, which lies between NGC 7319 and NGC 7318b, was first observed in radio continuum emission, and later in X-rays.  Observations strongly suggest the gas in the filament is heated by shocks driven into the intergroup medium by the "intruder" galaxy NGC 7318b \citep{gao00,xu03,xu05,osu09,igl12,rod14,kon14}. The galaxy is likely colliding from behind the group with a blue-shifted radial velocity of 800-1000 km s$^{-1}$ with respect to the majority of the gas in the group. The Quintet has also been the subject of several detailed numerical simulations \citep{ren10,hwa12,gen12}, which help to explain the HI and optical bridges and tails associated with the large Seyfert II galaxy NGC 7319, as well as the interaction of the intruder galaxy with the group \citep{mol97,sul01,sul95,wil02}.

{\it Spitzer} mid-IR spectroscopy of the filament revealed unusually bright pure--rotational emission-lines of molecular hydrogen emission (Appleton et al. 2006; Cluver et al. 2010; hereafter Paper 1) associated with the filament. Although the observations presented in Paper 1 could not spectrally resolve the emission or accurately measure the systemic velocity, a reanalysis of the Appleton et al. (2006) IRS higher resolution (R = 500) data was performed to provide some limited information
(See Appendix 1 of Paper 1).  Those results showed the gas to be kinematically broad (FWHM $\sim$ 800 km s$^{-1}$), and approximately centered at a velocity between that of the high-velocity intruder galaxy, and the velocity of the gas in the main body of the group. This suggested that the warm H$_2$ is highly turbulent post-shocked gas accelerated during the encounter.

The H$_2$ emission defines both the main north/south filament seen at other wavelengths, but also a  second bridge--like structure, which extends eastwards from the main shock towards the center of NGC 7319 (See Figure \ref{spitzerfoot}a). 
This possible bridge is also seen faintly in X-ray emission and extended H$\alpha$ observations \citep{tri03,xu03}. Based on the high values of the warm H$_2$/PAH, H$_2$/IR and H$_2$/X-ray luminosities (Paper 1), it was concluded that the main heating mechanism for the warm H$_2$ emission was the dissipation of mechanical energy in the shocks, channeled to small scales and low velocities by supersonic turbulence \citep[see modeling by][]{gui09,les13}. Recent observations of far-IR cooling lines ([\ion{C}{2}] and [\ion{O}{1}] and H$_2$O) with {\it Herschel} (Appleton et al. 2013) show that although these lines are boosted in the filament, the dominant cooling still occurs in the low-J pure--rotational lines of H$_2$.  

Early attempts at detecting CO~(1-0) emission from the Quintet appeared to only show emission from the brightest regions in NGC 7319, and in two extragalactic star forming regions \citep{gao00,smi01,lis02,pet05}. \citet{smi01} also detected faint broad emission from the direction of the intruder galaxy NGC7318b, but in retrospect, this may have been the first detection of CO from the main shocked filament which lies in the large NRAO 12-m beam.  

Much deeper single dish \citep{gui12} IRAM 30-m observations, using a broad-band receiver system, revealed the presence of extensive faint CO emission from the filament with large line-widths, and with as many as  three different velocity components at some positions. 
This multiplicity of velocity components \citep[also confirmed with optical spectroscopy of the faint ionized gas by][]{igl12,rod14,kon14}, implies a complex interplay between the
intruder galaxy's gas, and the pre- and post-shocked gas lying in the group. Unfortunately, although we expect the warm H$_2$ emission discussed in the current paper to primarily be post-shocked gas, we do not have the velocity resolution to make a detailed comparison with the previously published spectroscopy.
 
In Paper 1, we described the integrated properties of the warm pure-rotational mid-IR H$_2$ emission, including a comparison with the X-ray distribution, PAH bands and several fine-structure lines detected with {\it Spitzer}. In the current paper we explore for the first time the full 2-d excitation properties of the warm H$_2$ to derive physical properties within the gas. We also compare these data with observations taken of other gas phases, including [\ion{C}{2}] observations with {\it Herschel}, IRAM 30-m observations in the CO~(1-0) line, and new CO interferometer observations.  
We present three methods of modeling the H$_2$ excitation diagrams, including a simple two-temperature fit, a more general power-law technique, and finally multi-component shock-modeling. The shock-modeling allows us to quantify the mechanical energy dissipation in the gas caused by the intruder galaxy.    

The paper is organized as follows. \S2 of the paper describes the observations presented in the paper from { \it Spitzer} and {\it Herschel}, and \S3 summarizes the main observational results. 
\S4 describes three methods of modeling the H$_2$ excitation of the Quintet. \S5 shows the results of the modeling. \S6 explores the different estimates of the warm molecular mass and warm mass fraction using the various methods, and \S7 describes the spatial distribution of the derived properties, including kinetic energy disposition rates and shock properties both across and along the main filament. This includes a strong correlation between enhanced warm gas fraction and the kinematics of the gas \S7.1, as well as a discussion of additional heating from an AGN outflow from NGC 7319 \S7.2.  \S8, presents out conclusions, including the relevance of the observations for future far-IR studies of high-z galaxies and protogalaxies. For consistency with Paper 1, we assume a distance to the Quintet based on an assumed group systemic heliocentric velocity of 6600 km s$^{-1}$ of  94 Mpc  for H$_0$ = 70 km s$^{-1}$ Mpc$^{-1}$. 

\section{The Observations}
\subsection{Spitzer Observations}

Observations of the Quintet filament were made on 2008 January 11 and 2007 December 10 using  the short-low (SL with spectral resolution R = 60-127; SL2 covering 
$\lambda$5.2--7.7 $\mu$m, with sub--modules SL1, $\lambda$7.4--14.5$\mu$m ) and long-low (LL; R = 57-126, with sub--modules LL2  $\lambda$ = 14.0--21.3$\mu$m 
and  LL1, $\lambda$ = 19.5--38$\mu$m) modules of the {\it Spitzer} Infrared Spectrograph (IRS; Houck et al. 2004). A full description of the observations, data reductions 
and processing was presented in  Paper 1. The LL module (repeated separately for the 
sub-modules LL1 and 2) spectrally mapped an area
of $\sim$2.8 $\times$ 3.2 arcmin$^2$ in 21 steps of 8 arcsec (0.75 $\times$ the slit width), and was designed to cover whole radio/X-ray emitting filament. Figure \ref{spitzerspec} illustrates how 1-d spectra are built up to accomplish 2-d mapping. Similarly, the SL 
module (covering a slightly smaller area than LL) performed two partially overlapping scans perpendicular to the LL slit, in 2 $\times$ 23 steps of 2.8 arcsec (0.75 $\times$ slit width) 
for each sub-module.   After processing through the Spitzer Science Center S17 pipelines, all the individual frames were median combined, calibrated, corrected and assembled into 
2-dimensional data cubes using the CUBISM (Smith et al. 2007) software. This resulted in cubes for the SL  and LL mapped onto the sky with a pixel scale of 1.85 $\times$ 1.85  arcsec$^2$ and 5.1 $\times$ 5.1 arcsec$^2$ respectively. 

In order to ensure that we extracted spectra of the molecular hydrogen lines at a common spatial resolution, set by the longest-wavelength 28$\mu$m 0-0S(0) H$_2$ line, we convolved, at each 
wavelength, the individual layers of the SL1, SL2, and LL2 cubes (covering 5 to 21.3$\mu$m) to the scale of 7 arcsec (the FWHM of the 28 micron H$_2$ line) using a Gaussian kernel. It was not 
necessary to smooth the LL1 cube because the only line of interest in this paper is the 28.2$\mu$m line, and so this cube was left in its native form. This procedure has two positive effects. Firstly it 
ensured that our spectral extractions were not affected by resolution differences over the 5-28$\mu$m range of the H$_2$ lines (almost a factor of 5 in spatial resolution from 0-0S(5)6.1$\mu$m 
compared with 0-0S(0)28.2$\mu$m line). Secondly, the procedure improved the signal to noise of the shorter-wavelength data significantly (compared with Paper 1 which did not include smoothing), allowing us to produce high quality spectra over the full range of observed wavelengths.  The resulting data cubes were carefully checked for flux-conservation by extracting regions of the convolved cubes over large-scale features, and comparing the 
extracted fluxes with those obtained from the native resolution cube. The results were in very close agreement to a few $\%$.  

\begin{figure*}
\hspace{0.5 in}\centerline{\includegraphics[width=1.0\textwidth]{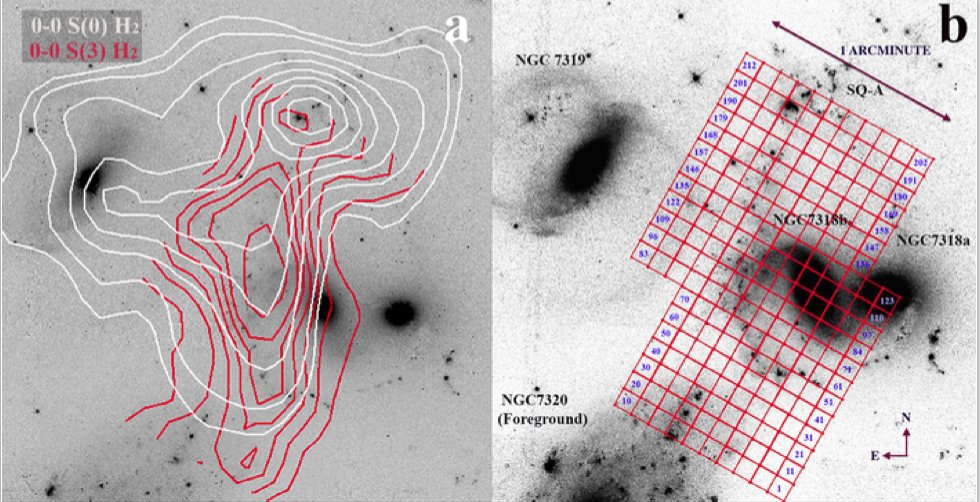}}
\caption{(a) {\it Spitzer} warm H$_2$ line maps of 0-0S(0)28.2$\mu$m (white contours) and 0-0S(3)9.7$\mu$m (red contours) superimposed on the HST F665N H$\alpha$ image of
Stephan's Quintet (See Paper 1). The 0-0S(0) map was made in the IRS long-low module and covers a larger area than the 0-0S(3) map, which was made in the short-low module.
White contours are 0.075, 0.1, 0.15, 0.2, 0.25, 0.3, 0.35, 0.4 MJy sr$^{-1}$, and red contours are 0.1, 0.2, 0.3, 0.4, 0.5, 0.6 and 0.7 MJy sr$^{-1}$, (b) Spectral extraction grid used in this paper to investigate
the excitation of the gas over the main H$_2$ filament, and only includes those areas common to both the long-low and short-low modules. The numbered extraction boxes
are 5.55 x 5.55 arcsecs$^2$ (2.5 x 2.5 kpc$^2$ at D = 94 Mpc) in area. The background image is again the F665N HST image. The overall extent of the warm H$_2$ emission is $\sim$50 x 35 kpc$^2$.  }
\label{spitzerfoot}
\end{figure*}

\begin{figure*}
\centerline{\includegraphics[width=7 in]{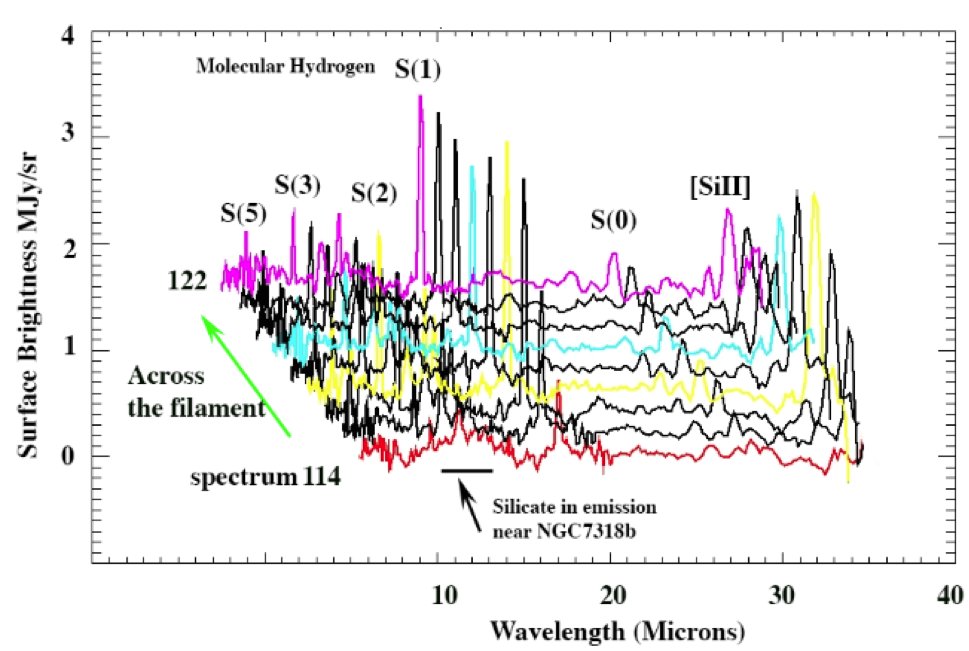}}
\caption{Schematically, we show a short run of nine (of the 212) IRS spectra from the Stephan's Quintet system (position 114 - 122 across the main shock; see  Figure~\ref{spitzerfoot}b). This demonstrates how the mapping was achieved. Examples of some sample spectra of regions of interest are shown in Appendix A (Figure \ref{figa1}a, b). The dominant emission is from the rotational lines of warm molecular hydrogen and [SiII]34.8$\mu$m. The red spectrum (114) is extracted close to the center of NGC 7318b, and shows silicates in emission around 10$\mu$m (see also Figure A1e). There is no significance to the colors of the spectra other than to make them more identifiable. Note the rapid rise in strength of the 0-0S(1) and higher rotational lines as one proceeds into the shock, and in this case across the high-excitation region of the so-called "Bridge" region (see text). }
\label{spitzerspec}
\end{figure*}

In this paper, we 
present extractions from the spectral maps over those regions that were mapped by both LL and SL. Spectra were extracted from 
the SL1, SL2,  LL1 and LL2 cubes, and then stitched together to form the full mid-IR spectra. In general, no scaling was needed in stitching the spectra together. The IRS spectral extractions were performed in CUBISM on the SL and LL cubes using the four corners of  3 x 3 native SL pixels to define the extraction areas over a continuous grid. This resulted in boxes 
of area 5.55 x 5.55 arcsec$^2$, which is close to half the width of the LL slit.  Figure \ref{spitzerfoot}b shows the extraction grid of 212 spectra superimposed on a continuum subtracted H$_{\alpha}$  HST (F665N) image of the Quintet.  We only sample regions 
that are well covered in both the SL and LL cubes, and so we have been conservative about the edges of the mapped area to ensure
good coverage at all wavelengths. The results of the extractions are presented in Table~A1.

\subsection{Herschel Observations} 
Observations were made  using the Herschel Space Observatory \citep{pil10}
PACS integral field spectrometer \citep{pog10} on 2011
December 7--8, respectively. In \citet{app13} we presented initial results and gas modeling of the [\ion{C}{2}] excitation
from these observations, showing that the [\ion{C}{2}] line flux is boosted by collisional heating from the warm H$_2$ gas. In the current paper we extend the analysis to include moment maps of the emission to explore the possible relationship between the [\ion{C}{2}] kinematics, and the heating of the warm H$_2$ gas.

\begin{figure}
\includegraphics[width=0.5\textwidth]{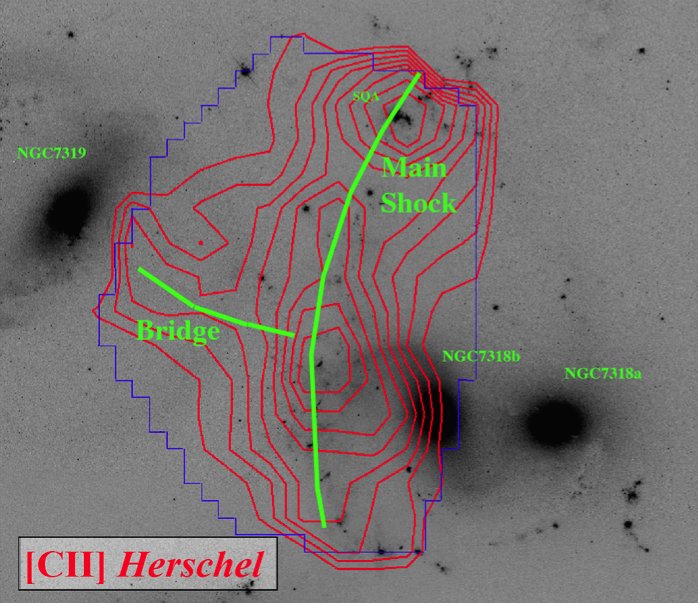}
\caption{ {\it Herschel} integrated map (contours) of the [\ion{C}{2}] emission in Stephan's Quintet obtained with the PACS spectrometer. The map is combined from three overlapping pointings of the IFU into one single image here. The blue outline marks the edge of the mapping area (contours extend slightly beyond these boundaries because of contour interpolation). The [\ion{C}{2}] emission  extends along the main N/S filament (including the extragalactic SF region known as the SQ--A), as well defining part of the bridge seen the {\it Spitzer} data, that extends towards the nucleus of the galaxy NGC 7319. Contour are 5, 7.5, 10, 12.5, 15, 17.5,19, 21, 25 in units of 1.2 $\times$ 10$^{-19}$ W m$^{-2}$ pix$^{-1}$.}
\label{Carbon2}
\end{figure}

The [\ion{C}{2}] 157.74$\mu$m observations were made in  
the first-order gratings using a short "range-scan"
mode covering the redshifted wavelength range 160.4--161.74$\mu$m (5050--7600 km s$^{-1}$ heliocentric)
with a velocity resolution of 235 km s$^{-1}$. The PACS integral field
unit (IFU) uses an image-slicer and reflective optics to project
5 $\times$ 5 spatial pixels (each 9.4 x 9.4 arcsecs$^2$ on the sky). 
Three separate "pointed mode" chop/nod observations were made
(3 arcmin chopper throw) with 4 hr of integration time per pointing
to cover the main parts of the Quintet filament \citep[see Figure \ref{spitzerfoot}b of][]{app13}. 
Unlike in the analysis presented in that paper, we combine all three pointings into
a single data cube using the software $specInterpolate$\footnote{See Herschel PACS User Manual}  which was not available at the time of that publication.  This software creates a single data cube combined from the three separate pointings, combining all the spaxel positions
using an interpolation scheme to ensure the best possible combination of the data given that the filament was not mapped
in a fully sampled way. The integrated [\ion{C}{2}] map of the Quintet filament is shown in Figure \ref{Carbon2} (left panel), and this was created by summing up the spectral flux in the projected cube over the range were emission is detected using the 
Herschel Interactive Processing Environment 12 (HIPE 12). 

\section{Observational Results}

\subsection{Mid-IR H$_2$ Emission and Spectra}
In Appleton et al. (2006) and in Paper~1, we presented strong evidence that the molecular hydrogen lines in the Quintet are excited by shocks over most of the filament. The lines are particularly strong
along the main filament (see also Figure \ref{spitzerfoot}a)
especially in the center.  Except from a few regions on the western side of the mapped region, there are very few places in the entire 212 extraction regions (Figure \ref{spitzerfoot}b) where no H$_2$ lines are detected.  The current paper concentrates on differences in excitation across the filament, including the region which extends to the east towards
the Seyfert galaxy NGC 7319, and westwards to include NGC 7318b and a. In Appendix~1, we present some examples of the extracted spectra to show the quality of these data, and the generally high signal to noise ratio of the molecular hydrogen line detections.  We note the detection of  Silicate emission peaking at approximately 10$\mu$m from spectra taken near the center of both NGC 7318a  and NGC~7318b (Figure A1e and A1h). Silicate emission is relatively rare, and its detection usually indicates the existence of a warm dust torus surrounding an AGN but seen from above \citep{li08,hat15}. We note that both galaxies show weak nuclear X-ray emission (O'Sullivan et al. 2009-Figure 2 of that paper), although NGC 7318a is brighter (with L(0.3-3keV) =  9.4 $\times$ 10$^{39}$ erg s$^{-1}$ for D = 94 Mpc; Trinchieri et al. 2005), and also contains a (0.95 mJy at 20 cm) nuclear radio source (Xu et al. 2003) with a spectral index of 0.62. It is therefore likely that both galaxies contain low-luminosity AGN.   

\section{Models of the H$_2$ Excitation}
Molecular hydrogen, being homonuclear, has no allowable dipole transitions, and so only weak quadrupole transitions of $\Delta$J transitions = $\pm$2 are allowed. H$_2$ excitation diagrams 
are a convenient way of exploring the distribution of level populations in H$_2$ molecules in galaxies (e. g. Rigopoulou et al. 2002, Roussel et al. 2007). This paper will compare three 
approaches to fitting the H$_2$ excitation data derived from the 2-dimensional IRS mapping of the Quintet. 
The first is a traditional method of fitting single or multiple temperature components to the excitation diagrams. This method is commonly
used in H$_2$ studies of galaxies. A second method is an extension of this idea,  assuming that most galaxy excitation diagrams can be
modeled as a single power--law distribution of temperatures. Both of these approaches assume that the gas is in thermal equilibrium. A third method assumes that the excitation of the warm H$_2$  is heated primarily by C- or J-shocks (or a combination of both) following the methodology of \citet{les13}. Arguments strongly in favor of shock excitation over other heating sources have been made
in previous works \citep[see particularly][]{app06,gui09,clu10,app13,les13}. In this case, LTE conditions are not assumed in the derived properties of the warm gas.   

\subsection{Fitting One or Two-temperature models to the excitation diagrams}

\begin{figure}
\includegraphics[width=0.5\textwidth]{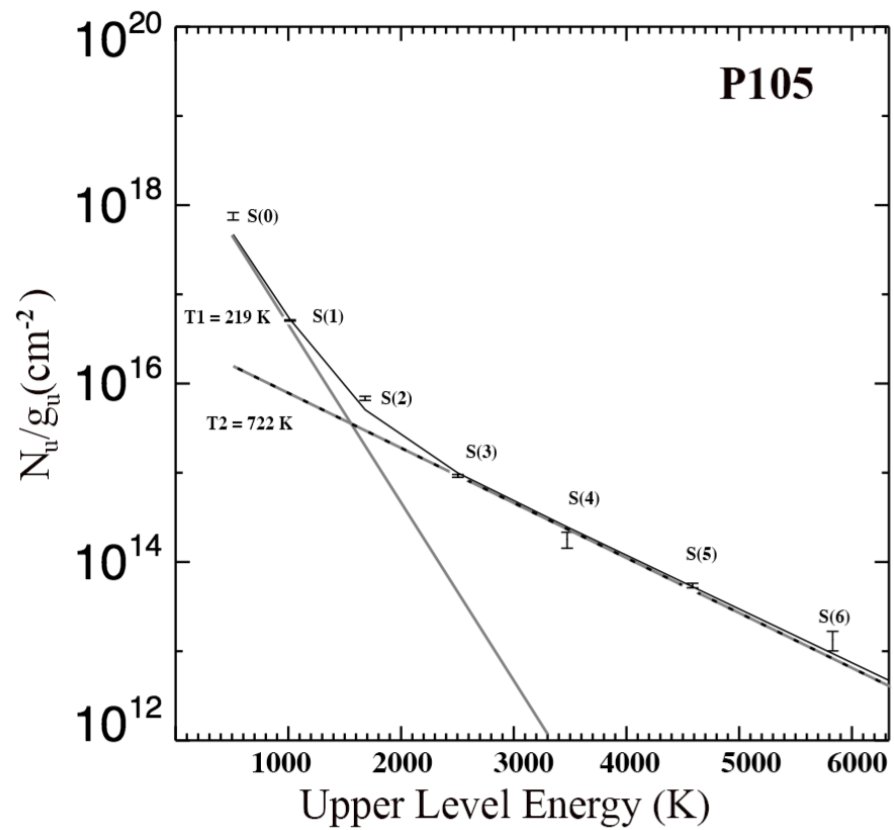}
\caption{A fit of the excitation diagram of warm molecular hydrogen with two temperatures for position 105 in the center of the the Quintet filament.The grey solid and dotted lines represent
the thermal components T$_1$ = 219K and T$_2$ = 722 K, which contribute to the final, black solid line fit. The values for each of the seven H$_2$ lines are shown with 1-sigma uncertainties. }
\label{2tempexcite}
\end{figure}

The transition we are considering are the lowest ground-state pure rotational states of H$_2$, e.~g.~0-0 S(0)$\lambda$28.22$\mu$m (corresponding to  J = 2-0),  0-0 S(1)$\lambda$17.04$\mu$m (corresponding to J = 3-1), 0-0 S(2)... 0-0 S(6)$\lambda$6.11$\mu$m. 
These transitions radiate in the wavelength range detectable with the {\it Spitzer} IRS\footnote{For the Quintet, transitions short ward of the 0-0 S(6) line were not detected.}  
Assuming the mid-IR lines are optically thin, the column density of the upper level of each pure rotational transition is measured observationally from the spectral line flux F of a given transition according to N$_u$ = 4$\pi$F/(h$\nu$A$\Omega$), 
where h is Planck's constant, $\nu$ is the frequency of the transition, A is the Einstein coefficient for the transition, and $\Omega$ is the solid angle of the observed region. In Local Thermodynamic Equilibrium (LTE),  
the upper level column density is related to both the excitation temperature T,  and the total column density N$_{tot}$ via,  N$_u$/g$_u$ =  (N$_{tot}$ exp [-E$_u$/kT])/Z(T), where E$_u$ is the energy of the upper level 
transition, k is Boltzmann's constant and Z(T) is the partition function\footnote{ Z(T) = 0.0247T/(1-exp(-6000/T), where T is in K; Herbst et al. (1996).}.  $g_u = (2S+1)(2J+1)$ is the statistical weight of the transition,
and $E_u$ is the energy of a particular upper level transition.  For $g_u$,  S is the spin number for a given J transition, where S= 0 for even J (called para-hydrogen) and S = 1 for odd J (ortho-hydrogen).   The H$_2$ 
excitation diagram
is usually presented as a plot of the log$_e$($N_u / g_u$) versus (E$_u  /k$), and is convenient because, for a single excitation temperature, the slope of a line fit to these points would be proportional to T$^{-1}$.    

It is generally assumed that, for the lower pure rotational H$_2$ transitions, the ortho and para-H$_2$ species should be in collisional equilibrium. As shown by Roussel et al. (2007) for H$_2$ densities $\gtrsim$ 10$^3$ cm$^{-3}$, 
most of the lower rotational transitions 
should be thermalized, and temperatures derived from fits to the ortho and para-H$_2$ transitions should yield consistent temperatures. Burton et al. (1992) showed that in collisional equilibrium, the ortho-to-para ratio (OPR) increase 
from about unity at T = 75K, to a constant value of 3 at T $>$ 250 K.  Since most of the temperatures we derive from the excitation fitting are in the range 120 $<$ T $<$ 600 K, the equilibrium values of OPR assumed are between 1.8 and 3. 
After normalizing for this factor, significant deviations from LTE would appear as an offset between the odd- and even H$_2$ transitions when plotted on an excitation diagram. We see no evidence for deviations from 
equilibrium in all our observations (See also Paper 1).

A single temperature fit is usually not found to be realistic, since the excitation diagrams are rarely consistent with a straight line, 
but are often curved.  In most cases a two temperature fit offers a  reasonable solution.   In the literature, this is a very common approach because it provides a first-approximation to the temperatures, column densities, and total H$_2$ masses derived 
from the observations. We will generalize this approach later in the paper to consider a power--law distribution of temperatures. However, we will begin with a simple one or
two-temperature fit to these data, under the assumption of LTE. This approach yields interesting insight into the excitation of the gas across this large-scale filament.

Single temperature solutions were found where there were not enough points to justify a two-temperature fit (22$\%$ of these data; see Table A2 for details), and two temperatures solutions were obtained for the other cases.  The process for fitting a two temperature LTE model to these data is iterative. First we create an initial excitation diagram with the lines fluxes (from Table~A1). Initially we assume an OPR of 3 to 
obtain a first approximation to the temperature, minimizing the chi-squared deviations for a two temperature model. Once the chi-squared values are minimized, we assume those temperatures as a first guess and adjust the OPR appropriate for  LTE from Burton et al. (1992) and re-run.  This process is repeated for several iterations until a best-fit value for the temperature 
and column density are determined. Figure \ref{2tempexcite} shows an examples of such a fit for a point near the center of the filament. The results of the fitting for the extraction regions of Figure \ref{spitzerfoot}b are presented in Table~A2.   For those minority of observations where the 0-0S(0) line was not detected, we provide  a range of possible temperatures for the gas. We use both a regular fit to the detected points, and a hard lower limit to the temperature, by treating the S(0) upper limit as a firm detection. This is important because the 0-0S(0) provides an anchor point for the lowest temperature gas, and effectively gives an upper limit to the H$_2$ mass which is dominated by the cold component. In cases where this was done, the values mapped closely to adjacent positions where S(0) was detected, implying that the lower temperature derived in this way has some validity.

\subsection{Power--law temperature model}

Considering the diverse range of heating environments in the galaxy ISM, the traditional method of fitting discrete temperature components to the H$_2$ excitation diagram is only an approximation, and may not be realistic in many cases--especially in the presence of shocks. Theoretical studies 
have demonstrated that H$_{2}$ line--cooling by molecules in shocks should follow a power--law temperature distribution \citep{Hollenbach79, Burton87}. \citet{Neufeld08} 
successfully used a power--law temperature distribution to fit the mid--infrared  H$_{2}$ rotational line fluxes of shocked regions in the supernova IC 443.

\begin{figure}
\centering
\includegraphics*[width=0.5\textwidth]{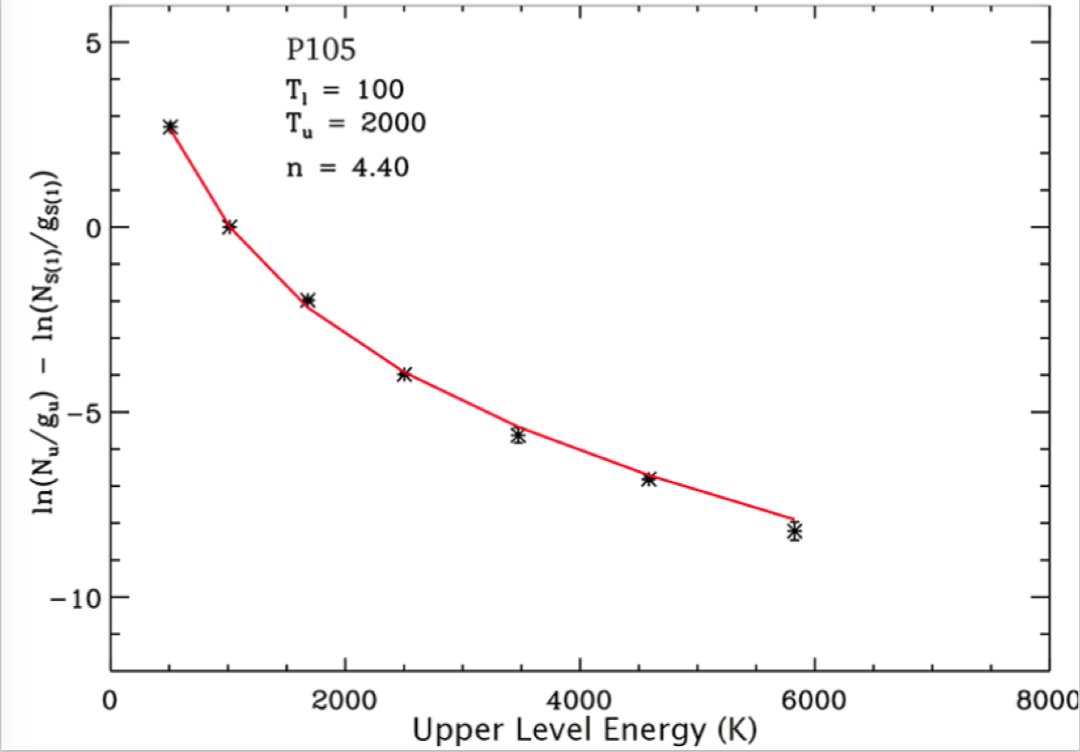}
\caption{Power--law model fit (red solid) to the observed H$_{2}$ line ratios (black points) in an excitation diagram for the square region numbered 105 of our mapped area (the same region as in Figure \ref{2tempexcite})}
\label{exd}
\end{figure}

We follow the approach of \citet{togi16} by assuming a continuous power--law temperature distribution for the H$_{2}$, and use this to fit the excitation diagrams and calculate the total H$_{2}$ gas mass in the ISM by extrapolating to lower temperatures. We assume that the column density of H$_{2}$ molecules are distributed by a power--law function with respect to temperature, dN $\propto$ T$^{-n}$ dT, where dN is the number of molecules in the temperature range T to T+dT. 
The model consists of three parameters, the upper and lower temperature bounds, and the power--law index, denoted by T$_{u}$, T$_{\ell}$, and n, respectively. T$_{\ell}$ can be thought of as the lowest (asymptotic) temperature found in the range of possible temperatures needed to explain the excitation of the gas. If  T$_{\ell}$ is well constrained (see later), and the single power--law approximation is valid,  molecular gas masses can be explored down to temperatures as low as  $\sim$50 K, significantly lower than those temperatures which excite the rotational transitions \citep[$>$ 80-100K;][]{togi16}.   Keeping the upper temperature, T$_{u}$, fixed at 2000 K and varying T$_{\ell}$ 
and n, we fit the H$_{2}$ excitation diagram for each square region in the grid of our mapped area of the Quintet. Figure \ref{exd} provides an example of a model fit (red solid line) to the observed H$_{2}$ line ratios (black points) in the excitation diagram of 
the square region 105, with $n$ = 4.4 in the temperature range 100--2000 K. This is the same region shown in the two-temperature fit of Figure \ref{2tempexcite}, and it is clear that the power--law fit more smoothly captures the change in shape of the excitation diagram than the two-temperature approach in this case. As discussed in Appendix 2, there are some regions were the power--law model fits less well, leading to a poorly determined value of T$_{\ell}$. 
\subsection{MHD Shock Modeling}

\begin{figure}
\centering
\includegraphics*[width=0.5\textwidth]{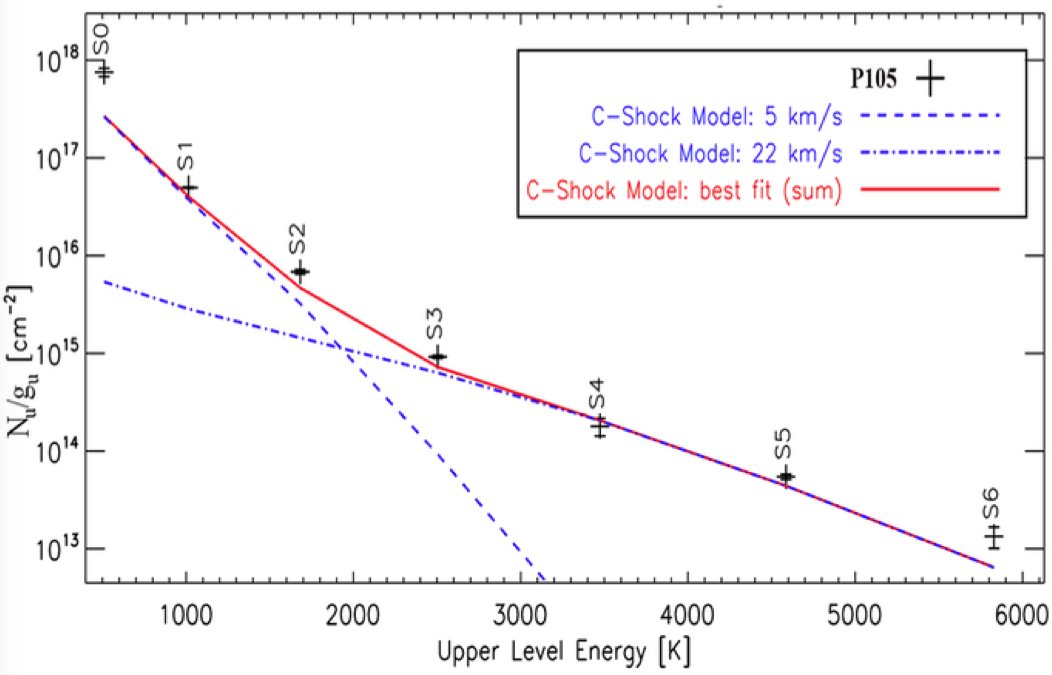}
\caption{H$_2$ excitation diagram fitted with a combination of two models at 5 and 22 km s$^{-1}$. The  pre-shock densities are 1$\times$10$^4$ and 1$\times$10$^3$ H cm$^{-3}$ respectively. }
\label{pgexcite}
\end{figure}

In a third approach to derive the H$_2$ physical parameters, and to place the heating of the gas on a more physical footing, we fit the observed H$_2$ line fluxes 
with the MHD shock models, using the approach presented in \citet{gui09}. We assume that the H$_2$ emission is powered by the dissipation of mechanical energy in  molecular gas at relatively low density ($n_{\rm H}$ = 10$^2$ - 10$^4$~cm$^{-3}$), as derived by \citet{gui09} and  \citet{app13}.
We use the grid of MHD shock models presented in \citet{les13}. The code computes the  populations of 150 rotation-vibration H$_2$ levels in parallel to the MHD equations.  The molecular gas is heated to a range of post-shock temperatures that depend on the shock velocity, the pre-shock density, and the intensity of the magnetic field (which is assumed to be perpendicular to the direction of the shock propagation). Our grid of shock models include a range of different shock speeds, from 3 to 35~km~s$^{-1}$, and we tested three pre-shock densities $n_{\rm H} = 10^2$, $10^3$, and $10^4$~cm$^{-3}$. The initial ortho-to-para ratio is set to 3, and the intensity of the pre-shock magnetic field is set to the square root of the preshock density (i.e. 30$\,\mu$G at $n_{\rm H} = 10^3$~cm$^{-3}$). The magnetic field corresponding to the minimum-energy equipartition is 10$\,\mu$G, derived from radio continuum observations averaged over large kpc-scales in the filament by \citet{xu03}. Therefore, the values used in the models are reasonable, given the much smaller scales that would be involved in the shocks \citep[see][]{gui09}.
The H$_2$ line fluxes are computed when the post-shock gas has cooled down to a temperature of 50~K and 120~K, chosen to compare our shock modeling results to single/two temperature or power-law fits. At a given pre-shock density, the shock velocity is the only parameter we allow to vary.  Figure~\ref{pgexcite} shows an example of an H$_2$ excitation diagram fitted with two shock models. 

We fit the H$_2$ excitation diagrams with a combination of two shock velocities, and two pre-shock densities. This provides a better fit to the data than fixing the same pre-shock density for both shocks, while keeping the number of free parameters to four.

In reality, we expect a distribution of shock velocities, but  \citet{les13} have shown that in the case of an preliminary analysis of the H$_2$ excitation in Stephan's Quintet (based on a single position in the filament), the favored probability distribution function of the shock velocities is a sum of two narrow Gaussian functions centered at two shock velocities. Following this approach, we simultaneously fit up to seven H$_2$ rotational lines for each position in the filament, and determine the best shock velocity combination which minimizes the reduced chi-square for each set of excitations.  
The H$_2$ masses at each point are derived by multiplying the gas cooling time (down to 100~K) by the gas mass flow (the mass of gas swept by the shock per unit time) required to match the H$_2$ line fluxes  \citep{gui09}. The total warm H$_2$ masses are obtained by summing these over the whole structure (See \S6.2 for comparison of the warm gas masses obtained by the various methods).  

\section{Model Results}

\subsection{One or Two-temperature analysis}
\begin{figure*}
\includegraphics[width=1.0\textwidth]{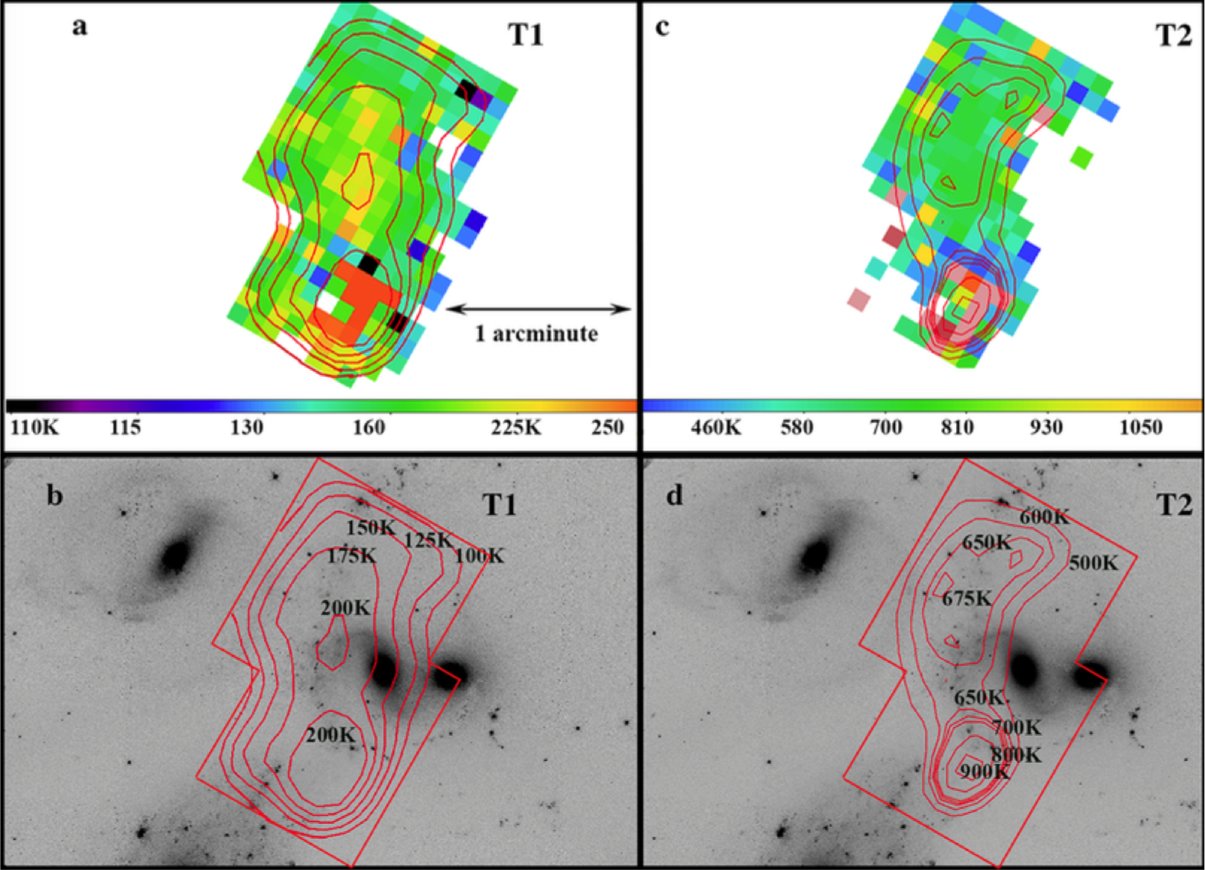}
\caption{Map of the temperature distribution of the warm H$_2$ gas derived from a two-temperature fit to the H$_2$ excitation diagrams at each sampled position. (a) lowest temperature (T$_1$) map image, with contours in K, (b) same map superimposed on the HST
F665N image of the Quintet, (c) the higher temperature component (T$_2$) an image with contours, (d) the T$_2$ contours superimposed on the F665N HST image.The contour representation includes some spatial averaging to show general trends. }
\label{temp2}
\end{figure*}

Figure \ref{temp2} shows the distributions of T$_1$ and T$_2$ temperatures across the Quintet. The T$_1$ (lower) temperature map Figure \ref{temp2}a and b, shows cooler gas in the northern part of the filament with a quite broad distribution, whereas the temperature peaks (T$_1$ $>$ 200 K) near the center and to the south of the main filament. The situation is similar for the T$_2$ component map (Figure \ref{temp2}c and d) where there are several peaks in the hottest component.   The hottest region in both the T$_1$ and T$_2$ maps is in the south of the filament. We will see that this is a general result of our modeling. The coolest T$_1$ and T$_2$ components lies near the extragalactic star formation site called SQ-A by \citet[][see label in Figure~\ref{spitzerfoot}b]{xu05}\footnote{We note that not all of SQ-A is covered by our observations}. The fact that the gas is hotter away from this star formation region supports the suggestion of Paper 1, based on an analysis of H$_2$/PAH ratios, that UV radiation from star formation is not the dominant heating mechanism in the filament. Another feature of the temperature maps shown in Figure \ref{temp2} is that  the warm H$_2$ component extends significantly in the direction of NGC 7319, and indeed H$_2$ emission is seen to the very edge of the eastern boundary of the region we are modeling. Unfortunately the spectroscopic diagnostic used in our current analysis (both the IRS LL and SL modules) do not extend
far enough to follow this distribution further to the east. However, we know from the maps using the LL module (see for example Figure \ref{spitzerfoot}a)
that the gas does extend all the way to NGC 7319. We will discuss the nature of this eastward extension in \S7.2.

\subsection{Power--law index analysis}
\begin{figure*}
\includegraphics[width=1.0\textwidth]{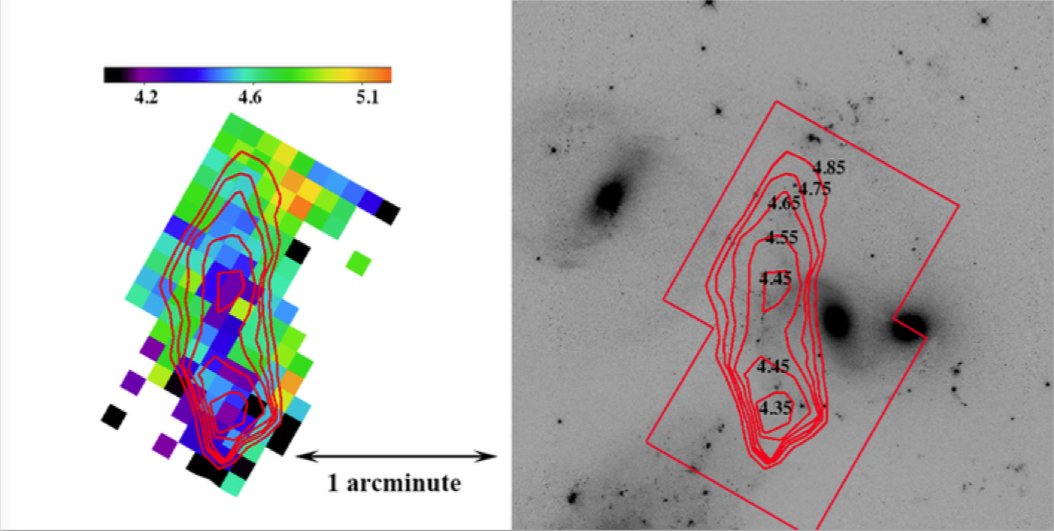}
\caption{A 2D color plot for the power--law index in our mapped regions of the Quintet. Smooth contours are overlaid on the color plot and the H$\alpha$ HST image from WFC3. The position of contours clearly mark the shock regions of the Quintet. The center of the shock structure has the lower power--law slope implying the presence of more excited warm molecular gas. The lowest fitted power-law values are in the center of the ridge line of the shock and to the south. The northern part of the filament has a higher power--law index implying lower excitation and generally lower temperatures over the range of temperatures sampled. This is similar to that seen in the two-temperature decomposition of Figure \ref{temp2}.}
\label{pli}
\end{figure*}

Figure \ref{pli}a shows a 2D color plot for power--law indices in the shock region of the Quintet with contours of equal power--law index superimposed. Detailed results are also presented in Table~A3. Figure  \ref{pli}b shows the same contours superimposed on the optical image of the galaxy. The contour levels map the main shock region of the filament very well, creating the impression of a curved structure which follows faint H$\alpha$ features seen on the HST image.   A large number of H$_{2}$ molecules are at high temperatures, leading to a flatter, or lower power--law index in warmest parts of the shocks because more power escapes from the high-J transitions. This is noticeable in the Figure where the power--law index shows a gradient from its highest values of 4.8-5 in the north of the filament, transitioning through the center of the filament (with a peak at 4.45), and reaching the lowest values (4.2-4.3)  in the south.  Because of the necessity in this fitting to include many lines, the modeling does not extend as far to the east as the two-temperature fitting, and so the extension in the direction of NGC 7319 is less obvious, although the fattening of the contours in the mid-shock region shows a bulge in the contours in the direction of the "bridge" marked in Figure \ref{Carbon2}. 

The values of the power--law index in Figure 8 are almost all lower than the mean found for normal galaxies of 4.84$\pm$0.61 (Togi \& Smith 2016), and they fall in the asymmetric tail in the distribution found in their study of SINGS galaxies. The tail is populated by SINGS galaxies containing  Seyfert  or LINERs nuclei. Indeed the only region that approximates to a normal power-law fit is the region near SQ-A, where the value for $n$ lies between 4.8 and  5.1. In the case of the centre of the shock-ridge, the values are consistently higher than any seen in the study of the SINGS sample. The shocks appear to be raising the temperature of the gas to much higher levels than are seen in more normal galactic environments. In \$7.1 we argue that this is a result of higher turbulent energy dissipation there, as measured by an increase in velocity dispersion in the diffuse gas. 

Another parameter in the power--law method
is the asymptotic value of T$_{\ell}$, the lowest extrapolated temperature obtained from the fitting process. In Appendix 2 we discuss how we calculate this, and how spatial variations in the H$_2$ excitation can affect how this lower limit is determined.  T$_{\ell}$ can affect how much mass in colder gas we estimate from the power--law method.  

\subsection{Shock models analysis}
 
A combination of two shock velocities is required to match the observed H$_2$ line fluxes, usually a very low velocity C-type shock (3-6~km~s$^{-1}$, and a stronger shock (20-35~km~s$^{-1}$), which is consistent with the models of \citet{gui09, les13} using limited preliminary data from this same survey. In order to fit best the excitation data for the 0-0S(0) and 0-0S(1) transitions, we found better results when allowing for two different pre-shock densities. The lower velocity shocks are associated with the higher density component.  We find that the pre-shock densities which provides the best fit to most of the H$_2$ line data is $n_{\rm H} = 10^3$~cm$^{-3}$ and a second component with $n_{\rm H} = 10^4$~cm$^{-3}$. On the outside of the ridge and in the bridge regions, pre-shock densities of $n_{\rm H} = 10^{2}$~cm$^{-3}$ and  $n_{\rm H} = 10^4$~cm$^{-3}$ are favored in our models. Although not unique, our fits  provide an estimate of the range of shock velocities and pre-shock densities needed to reproduce the H$_2$ excitation. This phase-space is well constrained when six or more H$_2$ lines are detected \citep{gui09, gui12}, which is the case for $\sim$25$\%$ of the spectra. 

Since we focus here on the properties of the H$_2$ excitation itself, this modeling do not attempt to simultaneously fit the H$_2$ and far-infrared 
([\ion{O}{1}] and [\ion{C}{2}]) line data. Joint fitting of [\ion{C}{2}] and H$_2$ lines in some small extraction regions, showed that the [\ion{C}{2}]-emitting gas is consistent with excitation by warm H$_2$ with densities of $n_{\rm H} \approx 10^{2-3}$~cm$^{-3}$ \citet{app13}. On the other hand, the discovery of pH$_2$O emission in the same paper requires that some components of the shocked gas require even higher densities $n_{\rm H}  > 10^5$~cm$^{-3}$. All of this suggests that a broader range of density structure is present in the gas that is not captured by the present models, which fit only the H$_2$ excitation.    

\begin{figure*}
\includegraphics*[width=1.0\textwidth]{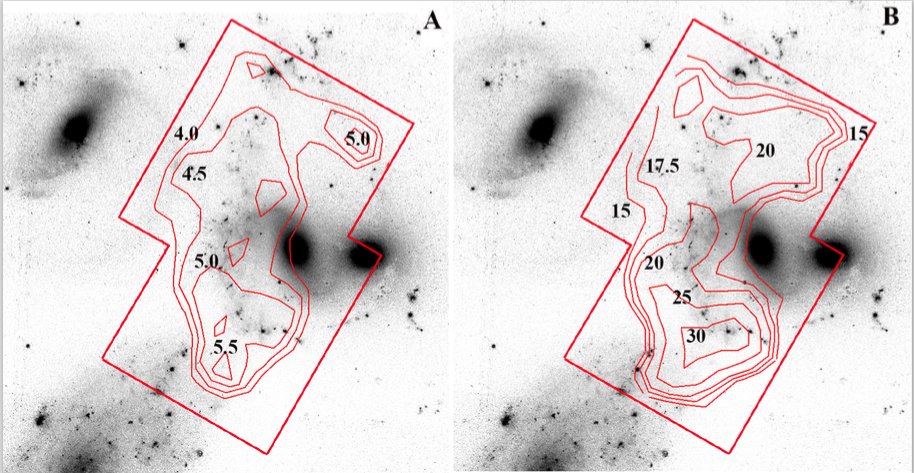}
\caption{The velocity field (in km~s$^{-1}$) of the (a) low and, (b) high shock components required to fit the H$_2$ excitation diagrams superimposed on the optical HST image of the Quintet. The general trend of increasing shock velocity from the north to the south is evident, as well as a bulge in both components to the east towards NGC 7319 (see text). We also note that higher velocity shocks are seen in the north-west between the arms of the tidal filaments from NGC 7318a and b. }
\label{shocks}
\end{figure*}

We performed the fitting analysis for all the spectral extractions in Figure~\ref{spitzerfoot}, and show maps of the two shock velocities in Figure~\ref{shocks}. Those maps clearly show the presence of stronger shocks in the center and southern part of the galaxy-wide shocked filament. The shocks are also stronger in the orthogonal direction of the filament, towards the bridge structure. 

\begin{deluxetable*}{lcccrc}{H}
\tabletypesize{\scriptsize}
\tablecaption{Warm Molecular Hydrogen (T $>$ 120 K) Mass and Warm Gas Fraction}
\label{tab:warm}
\tablewidth{0pt}
\tablehead{
\colhead {Method}  & \colhead{M(H$_2$)T$>$120K}  & \colhead {M(H$_2$)T$>$50K}  & \colhead{M(H$_2$)$_{CO}$\tablenotemark{a}} & \colhead {Warm (T$>$100K)}   & \colhead{Warm/Total ratio}\\
\colhead{} 	      & \colhead{(10$^8$M$_{\odot}$)} 	  &  \colhead{(10$^8$M$_{\odot}$)}  & \colhead{(10$^9$M$_{\odot}$)}  &   \colhead{Fraction}  & \colhead{ Method}\\
}
\startdata
2-Temp (All) & 11.2 & -- & 5.0 & 0.22-----------$>>$& CO obs\tablenotemark{a}\\
Powerlaw (All)     & 7.14 & 70.9 & -- & 0.1-----------$>>$  & p-l method\tablenotemark{b}\\
Shocks (All)  & 10.0 & 27.0 & -- &0.38-----------$>>$ & shock model\tablenotemark{c} \\
\hline
2-Temp T1 (only)  & 11.1 & -- & -- & -- & -- \\
2-Temp T2 (only)  & 0.12 & -- & -- & -- & -- \\
2-Temp (IRAM-R1\tablenotemark{e}) & 0.87\tablenotemark{d} & -- &1.3 & 0.07-----------$>>$ & CO obs\tablenotemark{a,e}  \\
2-Temp (IRAM-R2\tablenotemark{e}) & 0.69\tablenotemark{d} & -- & 1.3 & 0.05-----------$>>$ & CO obs\tablenotemark{a,e} \\
2-Temp (IRAM-R3\tablenotemark{e}) & 0.94\tablenotemark{d} & -- & 1.2 &  0.08-----------$>>$ & CO obs\tablenotemark{a,e} \\  
V$_{shock}$~$< $ 20 km s$^{-1}$ &  9.76 & 26.0 & -- & 0.38-----------$>>$ & shock model\tablenotemark{c}\\
V$_{shock}$~$\ge$ 20 km s$^{-1}$ & 0.40 & 1.0 & -- & 0.4-----------$>>$ & shock model\tablenotemark{c}\\
\enddata
\tablenotetext{a}{Using the total H$_2$ masses derived from Guillard et al (2012) based on CO (1-0) observations and an assumed N(H$_2$)/$I_{co}$of 2 $\times$ 10$^{20}$ cm$^{-2}$ [K km s$^{-1}$]$^{-1}$. This would include gas with a potentially lower temperature than 50K. }  
\tablenotetext{b}{Ratio of H$_2$ with T$>$ 120 K to the gas with T$>$ 50K using the power law method. This is the average value for the whole structure and may miss
colder gas that the method is not sensitive to. However, as discussed in the text and shown in Figure \ref{warmfrac}, this ratio can reach values $>$50$\%$ at the peak of the shock. }
\tablenotetext{c}{Total mass is evaluated from the shock model (see text).}
\tablenotetext{d}{Warm H$_2$ masses from the T$_1$/T$_2$ method were evaluated over the IRAM 30m CO~(1-0) beam. }
\tablenotetext{e}{A subset of positions along the shock ridge described in Guillard et al. (2013). These positions are dominated by warm gas with a high temperature and generally lower warm molecular masses than the average for the filament.}

\end{deluxetable*}

Note that the stronger shocks become J-shocks when their speed exceeds the critical velocity corresponding to the sonic point in the flow \citep{Leb02}. For a pre-shock density $n_{\rm H} = 10^2$~cm$^{-3}$ and a magnetic field induction $B = 10\,\mu$G transverse to the flow, the critical velocity is $V_{\rm crit} \approx 21$ km~s$^{-1}$. 
We note that we ran a grid of non-magnetic shock models (only J-shocks) and found very poor fits to the lower excitation lines (especially 0-0S(0)), stressing the importance of the magnetic field in softening the heating of the shocked gas, and enhancing the population of the lower excitation rotational levels with respect to higher ones. The shock model parameters, gas cooling times, mass flows, and warm H$_2$ masses are provided in Table~A4. 

\section{The Mass of Molecular gas}

\subsection{Warm Molecular Gas Mass and Warm Mass Fraction}

The three methods of fitting the excitation diagrams of the warm H$_2$ allow us to estimate the total warm molecular mass in the areas that we have sampled. For the two-temperature method, the fits for each regions can be summed to directly provide a total warm molecular mass for the whole mapped area.  For the power-law models, an important factor in determining the total warm molecular mass is the value of the asymptotic temperature T$_{\ell}$ (see \S 4.2) determined in the fitting process. This results in two kinds of behavior, gas in the center of the shock in which the bulk of the gas is warmer than 100-120K, and a second set of regions, mainly away from the main shock,  where the temperatures extend potentially down to 50K. The details of how we extrapolate this second kind of spectra is summarized in Appendix 2. Finally, the shock modeling also provides an alternative estimate of the amount of warm gas being processed through the shocks, since the models provide the mass flow rate into the shock and the cooling time for a given temperature. The total mass of warm gas is therefore the product of these two quantities for T   $>$ 120K.   The results are tabulated in Table~1. 

Given the different approaches used to measure the H$_2$ masses, the three methods agree well in terms of how much warm (T$>$ 120K) gas is present over the mapped area. For the two-component temperature fitting method, most of the mass is (as expected)
contained in the gas with the lowest temperature (T1) component (1.1 ${\times}$ 10$^9$M$_{\odot}$), with only  1$\%$ (0.12 ${\times}$ 10$^8$M$_{\odot}$) being contained in the warmer component, T2. The two-temperature method agrees well with the mass of warm (T$>$ 120 K) gas processed through shocks (1.0 ${\times}$ 10$^9$M$_{\odot}$). The power law method yields a slightly lower warm H$_2$ mass of 7.1 $\pm$1.5 $\times$ 10$^8$M$_{\odot}$ than both of the other methods, but we have been more restrictive in the area sampled by this method. Nevertheless, to within 30$\%$, all three methods are in agreement.
As Table 1 shows, the majority of the warm gas that is processed through the shocks lies in the lower-velocity shock component, whereas the mass of gas processed through the faster shocks is quite small.  

What can we say about the warm to total molecular fraction? \citet{gui12} estimated the total H$_2$ mass in the Quintet filament from IRAM 30-m observations  to be 5 $\times$ 10$^9$M$_{\odot}$ (assuming a value of  N(H$_2$)/$I_{co}$of 2 $\times$ 10$^{20}$ cm$^{-2}$ [K km s$^{-1}$]$^{-1}$ appropriate for the Galaxy). Adopting the two-temperature warm H$_2$ mass, we obtain a warm (T$>$120K) /total H$_2$ mass fraction of 0.22. However, since the IRAM beams did not cover the whole sitting region, the total mass of gas was only approximately estimated. We also include some estimates of the warm mass fraction obtained by the two-temperature method over the area of three individual IRAM pointings in Table 1. In those cases, where the warm H$_2$ mass is dominated by hotter gas, the ratio of warm to total H$_2$ mass drops to 0.1.

An alternative method for estimating to total molecular gas down to some lower temperature
has been presented by Togi and Smith (2016). The method relies on extrapolating the power--law index derived from the measured rotational lines
to much lower temperatures under the assumption that a single power-law index applies to all the gas. The method, which has a practical lower limit of T = 50 K (See Appendix 2 for more details), has some advantages over conventional methods. For example, its is based on direct emission from the H$_2$ molecules, and therefore does not need to rely on assumed relationships between other tracers (such as CO) and molecular hydrogen.    
The power--law method self-consistently can estimate the fraction of warm gas (say T $>$ 120 K) compared with the total H$_2$ mass found by extrapolating the power law down to cooler temperatures. Using this method the warm gas fraction is: 

\begin{equation}
\frac{M(>120 K)}{M_{total(T_{\ell})}} = \frac{\int_{120}^{T_{u}} T^{-n} dT}{\int_{T_{\ell}}^{T_{u}} T^{-n} dT}.
\end{equation}

where (see Appendix 2) T$_{\ell}$ has a practical lower limit of 50K.

\begin{figure}
\centering
\includegraphics[width=0.5\textwidth]{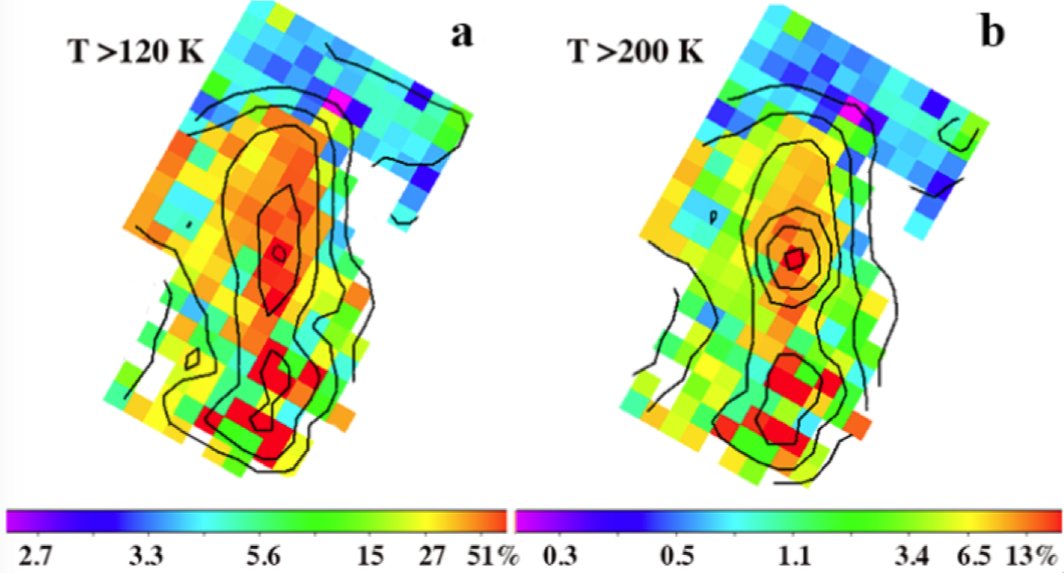}
\caption{The percentage of warm gas above a given temperature compared with that derived by extrapolating the power-law down to 50K (see text):  a) warm gas above 120K. Contour levels are 10, 20, 30, 40, 50 and 60$\%$,  and b) for hotter gas T $>$ 200K, with levels of 1, 2.5, 5,10,15 and 20$\%$.  The plot shows the warm gas fraction rises along the shock ridge as predicted by the models of \citet{gui09}. A smaller fraction is heated to higher temperatures near the shock center.  Note also that the region to the east (left) also contains a higher fraction of warm gas in the region of the so-called "AGN bridge". }
\label{warmfrac}
\end{figure}

Although the average warm fraction over the whole map, using this method, is  0.1 (Table~1), Figure~\ref{warmfrac} shows that the fraction is $>$ 0.5 in the center of the main shock (in a few places reaches unity), and is approximately 0.6 in the southern region, between NGC7318b and the foreground galaxy NGC 7320.  Given the uncertainty in extrapolating down to low temperatures using the power law method, and a similar large uncertainty and known scatter in the I$_{co}$/N(H$_2$) relationship used by \citet{gui12} in their estimate of the total mass of H$_2$, the results for the warm mass fraction using the various methods are consistent. 

\section{Properties across and along the main H$_2$ filament}

\begin{figure*}
\includegraphics*[width=1.0\textwidth]{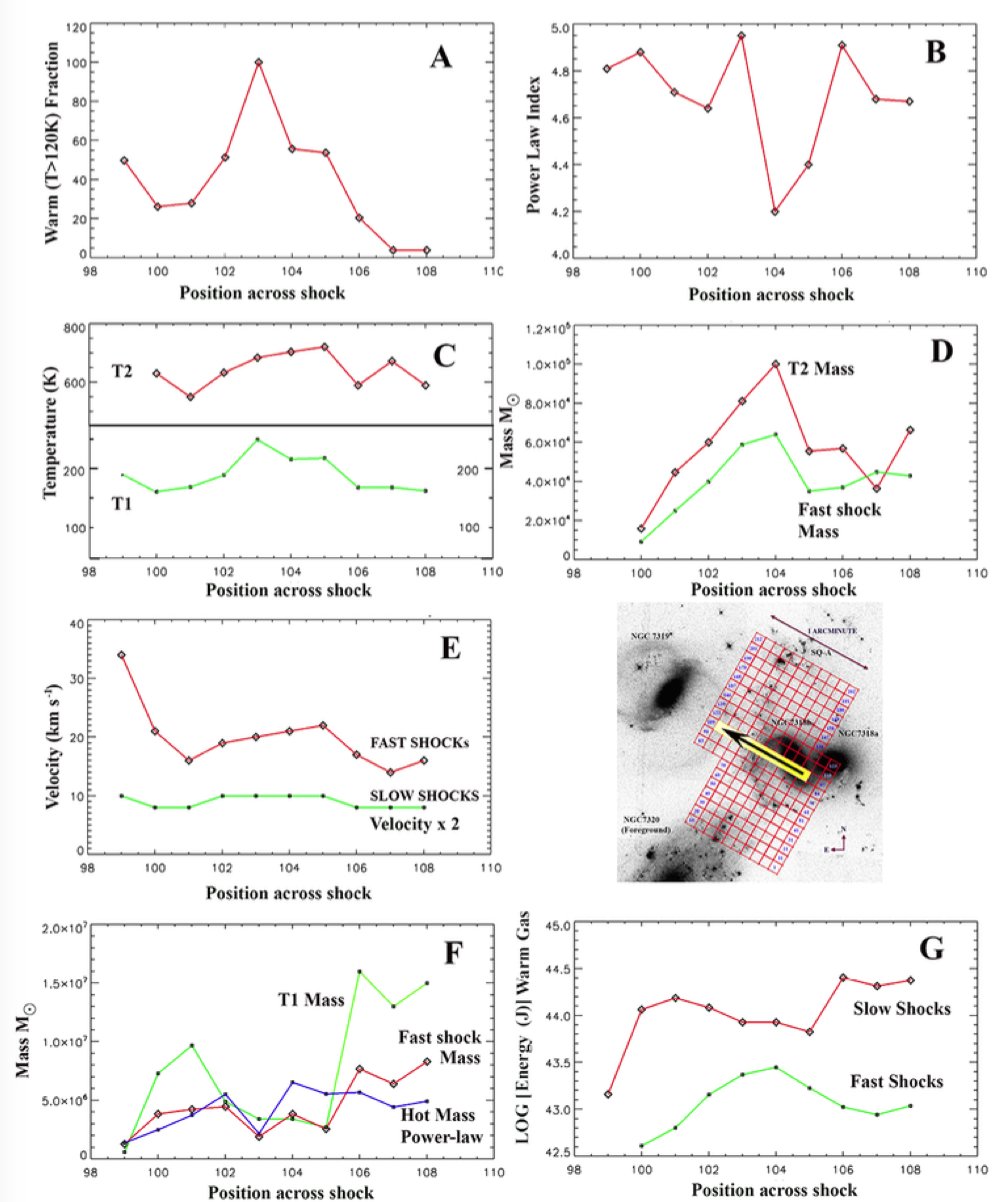}
\caption{Sectional slices across the H$_2$ filament following the arrow in the inset image for, (A) Warm H$_2$ fraction for T $>$ 120 K, (B), power-law index of H$_2$ (see text), (C) warm (T2, red line) and cooler (T1, green line) temperature distribution derived from two-temperature fits, (D) mass of warm H$_2$ associated with the warm (T2, red line), and fast shock (green line) component showing similar distributions, (E) velocities of the fast (red line)  and slower shocks (green line-velocities multiplied by 2 to show structure) needed to excite the H$_2$, (F)  gas mass associated with the cooler T1 component (green line), the fast shock component (red line) and the warm (T $>$ 120K) power-law mass (blue line), and (G) the kinetic energy in the warm (T $>$120 K) H$_2$ component associated with the fast (green line) and slow (red line) shocks. }
\label{slice1}
\end{figure*}

\begin{figure*}
\includegraphics*[width=1.0\textwidth]{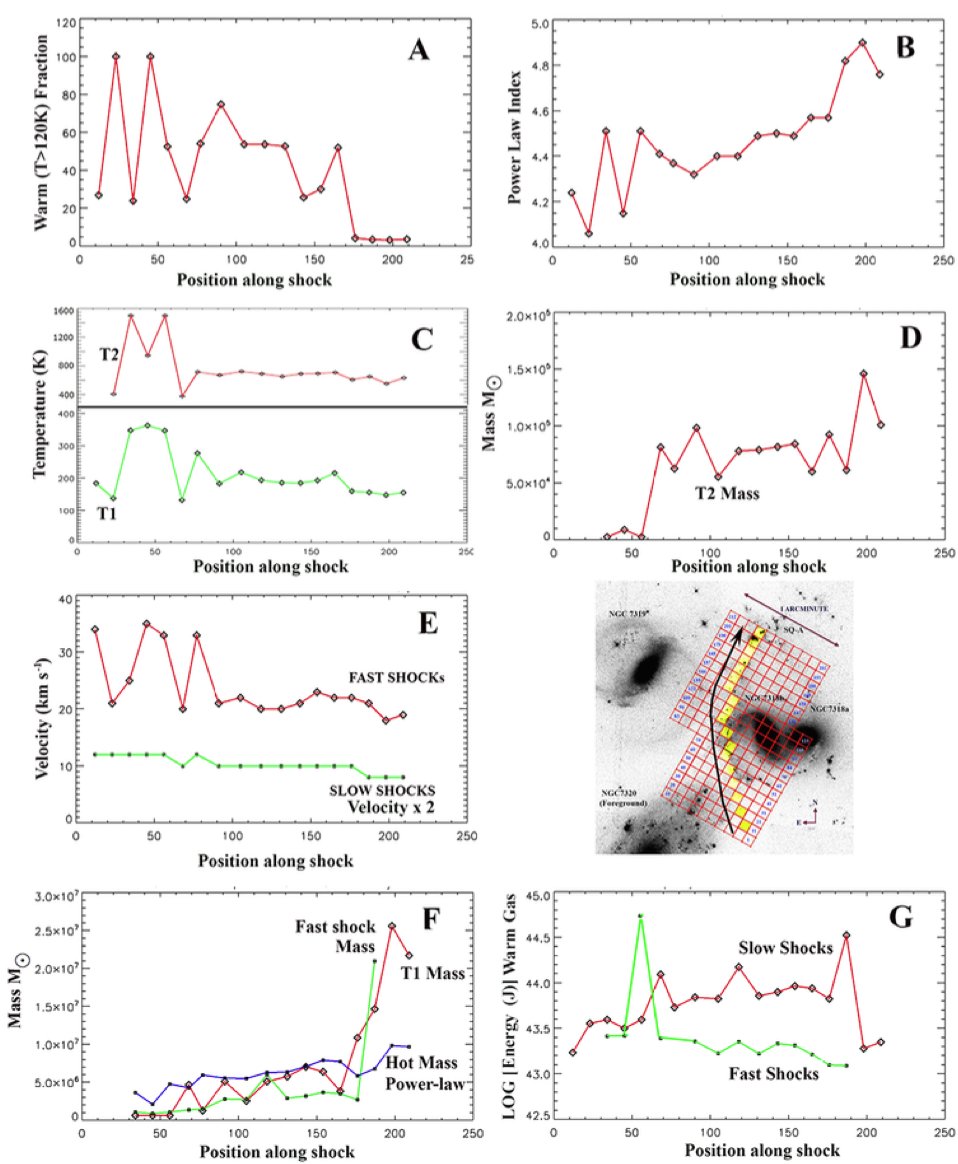}
\caption{Sectional slices (south through north) the H$_2$ filament following the arrow in the inset image for, (a) Warm H$_2$ fraction for T $>$ 120 K, (b), power-law index of H$_2$ (see text), (c) warm (T2, red line) and cooler (T1, green line) temperature distribution derived from two-temperature fits, (d) mass of warm H$_2$ associated with the warm (T2, red line), (e) velocities of the fast (red line)  and slower shocks (green line-velocities multiplied by 2 to show structure) needed to excite the H$_2$, (f)  gas mass associated with the cooler T1 component (red line), the fast shock component (green line) and the warm (T $>$ 120K) power-law mass (blue line), and (g) the kinetic energy in the warm (T $>$120 K) H$_2$ component associated with the fast (green line) and slow (red line) shocks. }
\label{slice2}
\end{figure*}

We present two "sectional slices" through the H$_2$ structure
both across the main shock (in Figure \ref{slice1}) from west to east, and  along the main shock from south to north (Figure \ref{slice2}). These slices help to illustrate some general 
properties of the gas in the main filament. The inset images (one per Figure) show the direction (arrow) and position of the slices we are considering. 

In Figures \ref{slice1}a and b, as we progress from west to east across the main shock, a sharp rise in the fraction of warm gas in the shock is observed, and at the same time a drop in
the power-law index at position 104, indicating high overall excitation there. In 
(Figure \ref{slice1}c), we see that the cooler gas temperatures (T$_1$) peaks first, followed by a peak in the warmer component (T$_2$) further to the east. In Figure \ref{slice1}d, we show both the molecular mass associated with the hotter (T$_2$) component, and the mass associated with the fast shocks (fast and slow shock velocities are shown in Figure \ref{slice1}e). Both track each other, and peak at the same position as the minimum in the power-law index.  This suggest that the warmer component essentially maps the fast shocks.  This is in contrast to Figure \ref{slice1} f, which shows the mass of gas associated with both the cooler component (T$_1$), and the low-velocity shock component. Both show a depression in the H$_2$ mass associated with the slower-shocks at the position of the peak, followed by a rise again beyond position 105. There is significantly more mass associated with the cooler components than the warmer, and the depression may be a real deficit of rotationally excited H$_2$ at the position of the maximum heating. The effect is still present (but less pronounced) in the plot of the total warm H$_2$ derived from the power-law fitting.

In Figure \ref{slice1}g we use the shock modeling to determine the kinetic energy dissipated by the shocks 
in the slow and fast shock cases. The kinetic energy deposited is  $0.5($\.{M} t$_{cool}$)V$_{shock}^2$ [J], (see Table~A4), where \.M is the rate of gas processing in M$_{\odot}$ yr$^{-1}$ (see Table A4).  The cooling time, t$_{cool}$, is very short (typically 10$^2$-10$^3$ yr). It can be seen that the fast shocks deposit more energy at the position 104, falling off on either side, whereas the slow shocks deposit more energy on either side of the filament, with a decrease at position 104.   Nevertheless, the slow shocks appear to deposit more total kinetic energy than the fast shocks.  Interestingly, the overall
kinetic energy deposited by the slow shocks continues to increase at positions beyond position 106.  This is the region that extends into the bridge region between the main shock and NGC 7319. 

In Figure \ref{slice2} we show a very similar set of diagnostic plots as the previous Figure, but this time emphasizing a line of points running from south to north along the main ridge of the filament (see inset image and arrow).  Here we see a general decrease in excitation from south to north in almost all the indicators. Figures \ref{slice2}a, c and  e show a general decrease in the warm mass fraction, the temperature (both T1 and T2) and model shock velocities (fast and slow) along the filament. On the other hand, the power-law index rises, and the mass associated with the gas also increase northwards for all components (Figure \ref{slice2} b, d and f).  Overall there is a relative flat kinetic energy deposition from the slow shocks from south to north, despite the decrease in overall shock velocities. There is also an indication that the fast--shock kinetic energy decreases from south to north. 

What could cause these global differences in warm gas properties and energy deposition on such a large scale? The basic picture that has been presented is that of a collision between
the intruder galaxy NGC 7318b and a pre-existing--almost linear--tidal HI filament \citep{wil02}. We have argued in previous papers (Appleton et al. 2006; Guillard et al. 2009; Paper 1)
that the warm H$_2$ can be understood in terms of the dissipation of mechanical energy through a turbulent cascade from a large-scale high-energy collision through supersonic turbulence
to small scales and higher densities. Guillard et al. (2009) presented a model in which two phases in the pre-shock gas were present in the HI filament before compression by the intruder galaxy. Slightly over--dense regions would be compressed by the high-speed shock from the intruder-galaxy, would lead to molecular cloud formation on grains which would survive the compression.  On the other hand, regions which were originally very under--dense would be shock-heated to millions of degrees (generating X-ray emission as observed), and the  grains would be destroyed (Guillard et al. 2009). This simple picture provides a working model of the system.  

The large extent of the H$_2$ in the north-south direction is consistent with idea of NGC 7318b colliding with a long tidal filament. The collision which is believed to be primarily towards
the observer from behind the group, may have occurred first in the northern part of the filament, and the point of impact may have progressed along the filament in a southerly direction. The fact that the northern part of the H$_2$ filament is cooler and heated by lower velocity shocks, implies that the energy has had more time to dissipate than in the south where, one might believe most of the current action is. 

The northern part of the filament also contains a larger mass of warm gas which radiates in the lower-J lines. This may be related to the structure of the initial target material before the collision of the intruder, implying a density gradient along that filament
which is now reflected in an increase in cooler gas. Given the short cooling time of the molecular gas (typically 1000 yr--see Table A4), we might expect this material to continue to cool rapidly. It is interesting that the northern part of the filament contains the extragalactic star forming region called "SQ-A". SQ-A is the only region of the whole filament that contains significant quantities of HI. Also, unlike the rest of the filament \citep{kon14}, this is one of the few regions of the  where significant current star formation is occurring (SFR = 1.5 M$_{\odot}$ yr$^{-1}$; Xu et al. 2003).  This may imply that this part of the filament may have cooled sufficiently for HI to re-form, and scattered star formation to be initiated. Indeed \citet{gal01} showed that SQ-A contains star clusters with a range of ages dating from the initial collision of NGC 7318a with the group (20Myr ago) to much more recent activity. Indeed recent observations support the earlier work (S. Gallagher, personal communication) which suggests that star formation activity in the main filament, though presently quite weak, is very recent, and may be increasing in intensity with time.   

Previous studies \citep{igl12,clu10} suggest a higher level of star formation not just in SQ-A, but also, to a lesser degree, in the southern part of the filament, near position 54 of Figure1b. Temperatures of the warm gas are a little higher there, but the study by Appleton et al. (2013; region  D of that paper) show that both H$_2$/PAH and [CII]/PAH ratios are inconsistent with PDRs dominating the H$_2$ heating there.
Indeed, although the H$_2$ is warmer in that region, the heating is not confined to the \ion{H}{2} regions, but is more extensive, suggesting that the extra heating is not caused by
star formation. 

The structure of the main filament is quite different in the east-west direction--as we have shown. Perhaps surprising is the extent of the gas in the easterly direction where
it continues to the edge of our grid. Long-low observations with IRS extended further to the east than that of the Short-low observations, and so we know that the warm H$_2$ continues eastwards.  
In Paper 1 we suggested that the eastward extent of the emission was part of a "bridge" to NGC 7319 (See Figure 1a). However such a bridge would have to be quite broad since
we detect warm H$_2$ as far north as position 212 (estimated temperatures of T1= 143K and T2 = 463K). This suggests that a large volume of gas is caught-up in the shock-heating. 

It is worth considering the implications of the kinetic energy deposition in the warm gas. Both across and along the main H$_2$ filament (Main Shock of Figure 1a), within each element the kinetic energy deposited every 500-1000 years (the gas cooling time see Table A4) is between 10$^{43.5-44.5}$J per observational element (6.2 square kpc), and summing over the whole observed region is 6.3 $\times$ 10$^{46}$ J (6.3 $\times$ 10$^{53}$ erg), and a total energy flow through the shocks in the warm gas of 4.4  $\times$ 10$^{36}$ W (4.4 $\times$ 10$^{43}$ erg s$^{-1}$).
The total H$_2$ line luminosity from the whole region (summing over the detected 0-0S(0)-S(5) lines) is 1.3$\times$ 10$^{35}$ W (1.3 $\times$ 10$^{42}$ erg s$^{-1}$), and so the efficiency of conversion of kinetic energy in the shocks into warm H$_2$ line luminosity is $\sim$3$\%$. 

\subsection{Connection between Large-scale Gas Kinematic and Regions of Maximal Heating}

An interesting question is whether the regions of maximal heating in the shocks (as measured by for example the distribution of warm H$_2$ gas fraction) bears any connection to the large-scale dynamics of
the Quintet system?  NGC 7318b is colliding with intergroup gas at about 1000 km s$^{-1}$ and Guillard et al. (2009) proposed that, through a turbulent cascade, energy is being transferred to smaller scales and low velocities where a significant fraction of the energy escapes through rotational H$_2$ and far-IR fine structure emission (Appleton et al. 2013). 

The shock modeling refers to the micro-scale (pc scale or smaller) shocks which represent some of the smallest scales in a turbulent cascade. The large-scale kinematically-broad lines measured in the ionized and diffuse molecular gas  likely represents the largest velocities and scales in the turbulent cascade. In our picture of the turbulent cascade, we consider the broader lines as being composed of a multitude of molecular shocks and turbulent eddies of narrow width ($<$ 30 km/s) spread over the full range of observed large-scale stirring (caused by the larger scale collision of NGC 7318a).

\begin{figure}
\includegraphics[width=0.5\textwidth]{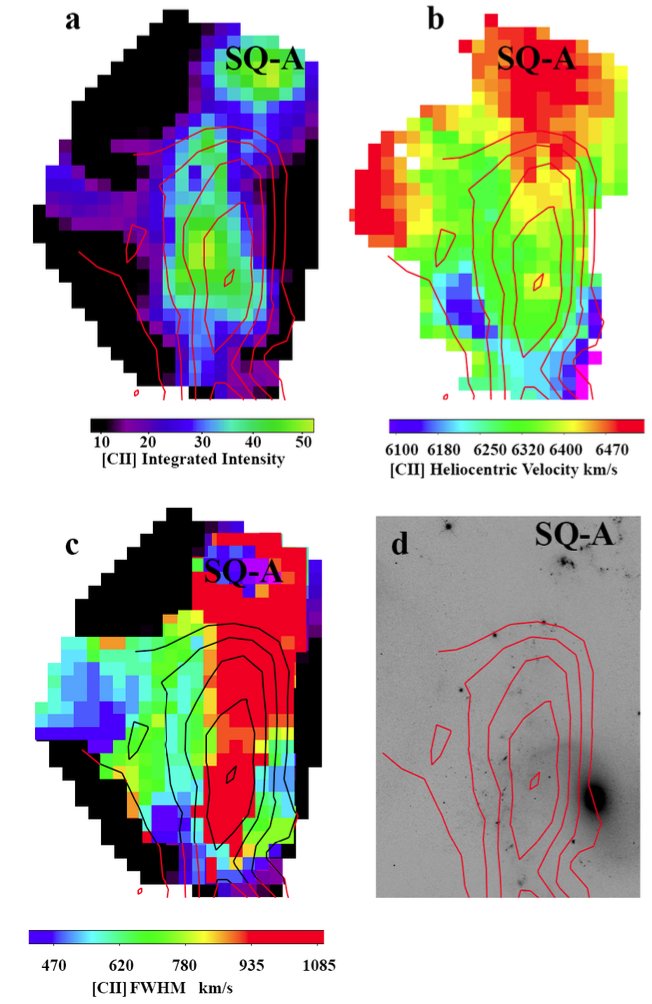}
\caption{a) The warm (T $>$ 120 K) mass fraction (contours in steps of 10$\%$ starting at 10$\%$) compared with the moment maps for the [\ion{C}{2}] emission from {\it Herschel},  (a) [\ion{C}{2}] integrated intensity (arbitrary units),  (b) [\ion{C}{2}] heliocentric mean velocity of the emission (km s$^{-1}$), (c) [\ion{C}{2}] line-width (FWHM) (km s$^{-1}$), and (d) compared with the HST F665N visible image. SQ-A, which contains recent star formation is also shown.  The largest warm H$_2$ gas fraction lies to the west of the main ridge of [\ion{C}{2}] emission, but correlates well with the regions of maximum velocity dispersion. This suggests the [\ion{C}{2}] emission maps the leading edge of the main shock, and gas to the west is where the kinetic energy is being deposited, increasing the warm H$_2$ gas fraction in the post-shocked turbulent regions. }
\label{CIIionized}
\end{figure}

If energy is indeed being fed down from large to small scales,  then those regions of the gas which show large global velocity dispersions should correlate with the regions of maximal heating. Since the velocity resolution of the IRS was not sufficient to map the kinematics of the H$_2$, we turn to the [\ion{C}{2}] emission as a tracer of the diffuse warm molecular gas. 
It is likely that the warm H$_2$ is the main collisional partner for the diffuse ionized Carbon, leading to the excitation of  [\ion{C}{2}]158$\mu$m transition in the shocked region (Appleton et al. 2013). We show in Figure \ref{CIIionized} a, b and c, contours of the warm molecular fraction over--plotted on moment maps (intensity, radial velocity map, and line--width) for the [\ion{C}{2}] emission over the areas that they overlap. Figure \ref{CIIionized}d shows the same contours on the optical image, for reference. From this Figure, it is clear that the regions of highest warm fraction correlates most strongly with those regions with the largest ($>$ 1000km s$^{-1}$) [\ion{C}{2}] line--width. This implies that the high levels of heating of the warm H$_2$ is strongly correlated with more turbulent line widths (large-scale stirring), as expected by the model of Guillard et al. (2009). We note that the peak in the [\ion{C}{2}] emission ridge is offset to the east (Figure \ref{CIIionized}a) from the region of most intense H$_2$ heating, suggesting that the [\ion{C}{2}] emission peaks ahead of the region of maximum kinetic energy dissipation. This would be consistent with the idea that the main shock has a component of motion to the east. It is also consistent with Figure \ref{slice1} which shows hotter
gas temperatures, and faster shocks to the east of the main ridge of the peak in the warm gas fraction (Figure \ref{slice1}d and e compared with \ref{slice1}a). Given that the main disturbance driven into the intergalactic gas is a three-dimensional surface, the actual motions in the gas are likely to be quite complex and not easily interpreted without
a realistic numerical simulation. None of the existing simulations mentioned in the introduction provide enough detail to explain all the facets of our observations.  

\subsection{Excitation of gas in the "Bridge"}

Most of the discussion so far has concentrated on the main N/S filament associated with the main shock of Figure \ref{Carbon2}. We turn now to discuss
the emission to the east of the main shock, especially the "bridge" marked in Figure \ref{Carbon2}. The bridge has been previously detected in X-ray emission and diffuse H$\alpha$ emission \citep{tri03,xu03}, and warm and cold molecular gas (Paper 1; Guillard et al. 2012). It was not detected in radio continuum emission, unlike the main shock. 

\begin{figure}[h]
\includegraphics[width=0.5\textwidth]{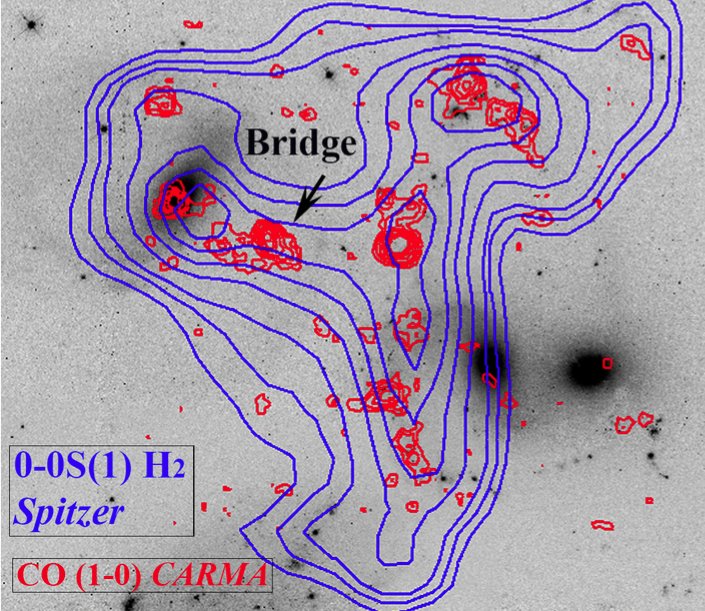}
\caption{Red contours showing the CARMA CO~(1-0) integrated emission (Guillard, Appleton \& Alatalo, in preparation) superimposed on the HST F665N image (see companion paper of Guillard~et~al.~2016 for full discussion of CO emission).  Blue contours show the integrated surface density of the 0-0S(1)17.08$\mu$m emission covering most of the filament and NGC 7319 from Paper 1. Note the clumps of CO emission define both the main shock, but also are scattered along the bridge to NGC 7319 }
\label{COmap}
\end{figure}

As shown in Figure \ref{CIIionized}c,  by comparison with normal galaxies, even the bridge region has excessively large [\ion{C}{2}] line-widths of $\sim$500 km s$^{-1}$, which may explain why so much of the H$_2$ is excited over such a large area. The bridge structure has its own distinct [\ion{C}{2}] kinematics, as can be seen in Figure \ref{CIIionized}b, where even in this limited coverage of the full bridge, a strong velocity gradient is seen of 300 km/s over a linear scale of $\sim$15 kpc, with the highest velocities seen closer to NGC 7319. The IRAM 30-m CO~(1-0) and (2-1) observations of the bridge region by Guillard et al. (2012) shows that this region contains multiple broad CO-emitting components with gas at velocities intermediate between those in the extreme north of the group and those in the south (this can been seen also in the [\ion{C}{2}] velocity field of Figure \ref{CIIionized}b). Figure \ref{COmap} shows new higher resolution observations of some brighter CO condensations within the bridge made with the CARMA interferometer superimposed on the 0-0S(1) Spitzer map. These data (which will be discussed in much more detail in a companion paper; Guillard et al. in preparation) imply a coherent string of multi-phase gas emission extending from the Seyfert nucleus of NGC 7319 into the intra-group medium. This raises the possibility that there is an atomic and molecular outflow associated with the nucleus of NGC 7319.  A full discussion of this possibility is beyond the scope of the present paper, since both the warm H$_2$ diagnostics and the [\ion{C}{2}] emission do not extend far enough to explore a possible jet-heating scenario. 

We have already shown in \S7 that as one progresses further east from the main shock, the excitation of the cooler component of the warm H$_2$ gas remains high.  Figure \ref{slice1}f and g show that in the region of the bridge (Positions 106, 107 and 108), the dominant heating is from slower shocks associated with cooler (T = 150 K) gas which contains a significant mass of H$_2$. As Figure \ref{slice1}a shows, this is also a region with a much lower warm H$_2$ fraction based on the power-law analysis, suggesting that the balance has shifted to a large reservoir of cooler, but still excited gas in the bridge.  We might speculate that the source of heating of this 150 K gas is kinetic energy injected into
the medium by an AGN outflow from NGC 7319. This process is evidently less efficient at heating the gas that the collisional-driven shock from NGC 7318b. More work will be needed to explore this possibility in a future paper (Guillard et al. 2016). 

\section{Conclusions}

We have analyzed the two-dimensional excitation distribution of warm molecular hydrogen based on 212 individually extracted mid-IR spectra from {\it Spitzer} IRS observations of the Stephan's Quintet intergalactic filament. The analysis was conducted on those regions with common IRS short-low and Long-low spectral coverage to allow detailed fitting of the molecular excitation. Our main conclusions are:  

\begin{itemize}

\item{Using simple linear fits to the H$_2$ excitation diagrams, we have mapped for the first time the temperature and mass distribution of warm H$_2$ across the intergalactic gas in Stephan's Quintet. The mapped region extends over an area of at least 35 x 55 kpc$^2$, and includes the area dominated by shocked gas ahead of the intruder galaxy NGC 7318b. The warmest gas (600-1000 K) is found in the center and to the south of the main filament, with an extension of warm gas extending towards the Seyfert galaxy NGC 7319.  The extragalactic star forming region called "SQ-A", at the northern tip of the main filament, contains the coolest of the warm H$_2$.  A total of 1.1 $\times$ 10$^9$ M$_{\odot}$ of warm (T $>$ 120 K) H$_2$ is detected across the entire mapped region, with the largest contribution coming from the cooler (T $\sim$ 120-150K) extended regions on either side of the narrow highly shocked region, and from SQ-A.}

\item{We also fit the H$_2$ excitation diagrams with a power-law distribution of H$_2$ column density with temperature dN$_{H2}$ $\propto$ T$^{-n}$ dT. The power-law indices range from n = 5 in the north (lower excitation) to lower values (higher excitation) of n=4.2-4.3 in the southern filament. Except in the north, the measured power-law indices are much lower than those seen in normal galaxies, reflecting the high excitation likely resulting from shock heating. The distribution of the lowest power-law values defines a curved band lying to the East and ahead of the intruder galaxy NGC 7318b, as expected if the galaxy is driving a large-scale disturbance into the group-wide gas. The power-law method, self-consistently, can predict the mass of cold gas at each point in the map. This analysis shows that along the crest of the main filament, where shocks are expected to be the most intense, the warm gas (T$>$120 K) fraction is between 60--100$\%$. } 

\item{We have found that the warm H$_2$ fraction correlates strongly with the large-scale kinematics of the gas (as measured by the [\ion{C}{2}] emission). Where the FWHM of the velocity field is $>$ 800 km s$^{-1}$, the warm H$_2$ is heated the most strongly. This provides compelling evidence that large-scale turbulent stirring of the diffuse gas caused by the collision of NGC 7318b with the group is able to influence and heat gas on the small scales and low-velocities needed to excite the rotational H$_2$ lines.}  

\item{We modeled the H$_2$ excitation with a molecular shock modeling code \citep{les13} which computes, along with other common fine-structure lines, the populations of 150 rotation-vibration H$_2$ levels in parallel to the MHD equations. Although the models are not unique, the best results were achieved with shocks propagating in a molecular medium containing gas with two different pre-shock densities.  A system of slow magnetic C-shocks (4-5 km s$^{-1}$) and faster mainly J-shocks (up to 25 km s$^{-1}$) were required to explain the observed emission. The models support the idea that the high speed disturbance caused by the intruder (at 1000 km s$^{-1}$) has been degraded into millions of slower molecular shocks through the effects supersonic turbulence.}

\item{Our results provide a picture of the dissipation of kinetic energy through compressive turbulence (shocks) both along the length of the main filament (north--south) and across it, mainly East--West. Gas temperatures rise, the power-law index flattens and the velocity of the shocks needed to excite the gas increases as we move from north to south along the main filament. A similar phenomenon is seen across the shock in the east-west direction, where the hottest most intense shocks are seen at the peak of the H$_2$ surface density, with a slow drop--off on either side. However, although brighter at the peak, the estimated total surface density in warm H$_2$ actually decreases with this increasing temperature. This may reflect the density structure in the initial pre-shock material that was overrun by the main disturbance from the intruding galaxy, or may be a result of molecules being increasingly dissociated in the most excited region of the structure.} 

\item{Our analysis shows that the H$_2$ gas continues to be warm and excited far to the east of the main filament, in the region of the bridge that connects to NGC 7319. Some of this may be related to a known collimated radio jet. Our [\ion{C}{2}] observations show an extension of diffuse gas towards the nucleus of this galaxy associated with a kinematically shifting component, but the observations do not extend far enough to conclusively associate this with an outflow. However, our CO~(1-0) map from CARMA  covers a larger area, and shows a string of dense H$_2$ clumps extending from NGC 7319 in a similar region to the bridge, suggesting a connection with the center of NGC 7319.  If the AGN in NGC 7319 is somehow exciting warm molecular hydrogen in that direction, it must have a broad effect, since the warm H$_2$ gas is extended on a large scale, and is generally cooler than the  intra-group gas associated with the main filament.}

\end{itemize}

The observations of shock-heated gas in Stephan's Quintet emphasizes the wealth of information that can be obtained from rest-frame mid-IR molecular lines, and may point the way towards future studies of turbulent gas a higher redshift with {\it JWST}, {\it SPICA} and other possible future facilities. Systems with similar or more powerful rotational H$_2$ lines were discovered towards the end of the {\it Spitzer} cryogenic mission. These include a wide range of environments from nearby shock-excited systems and groups \citep{app06,pet12,clu13, ala14}, low-redshifts radio galaxies \citep{ogl07,ogl10,nes10,gui12b,ogl14,lan15} and central cluster galaxies \citep{ega06,don11}. It is interesting that the most powerful H$_2$ emission found to date has been from the Spiderweb protocluster at z = 2.15 \citep{ogl12}, and from stacked IRS spectra of z $\sim$ 2 galaxies \citep{fio10}. Although the emission mechanisms for the high-z gas is not yet clear, the IRS points the way towards potentially important discoveries in the future \citep{app09,gui15}. 

One question that is raised by the current observations is that of the dominance of molecular hydrogen cooling versus IR fine-structure lines.  For example, in the pioneering work of
Santoro \& and Shull (2006), these authors describe how the initial formation of the first galaxy-sized clumps of gas is dominated by
cooling from mid- and far-IR rest-frame fine-struture lines (principally [SiII]34.8$\mu$m, [\ion{C}{2}]157.7$\mu$m and [\ion{O}{1}]63.2/145.5$\mu$m), once the metallicity increases above a critical value (typically Z $>$ 10$^{-3}$ solar. In such a case, the initial collapse of the first clouds would radiate mainly in the metal fine structure lines, with the mid-IR H$_2$ lines playing a minor role in the cooling. However, once the collapse is initiated, many complex processes, including shocks and turbulence could rapidly change the conditions in the gas. Our purely molecular models of Stephan's Quintet's shock (Lesaffre et al. 2013, Guillard et al. 2009, and the current paper) predict comparable H$_2$ and [\ion{C}{2}] line luminosities, as observed (Appleton et al. 2013), even in regions of much higher metallicity than those expected in primordial clouds. A lesson from the Stephan's Quintet system is that the collision of NGC 7318b with a mainly HI tidal arm has resulted in an almost total transformation of the HI into H$_2$ in 10-20 Myr (See Guillard et al. 2009).  We have also shown here that most of the kinetic energy is bound up in the coolest gas heated by low-velocity  (V $\sim$4-5 km s$^{-1}$) shocks.  If low-velocity shocks form at the end-point of a turbulent cascade in gas accreting within dark-matter halos, the H$_2$ lines may again becoming important in the cooling of the gas, perhaps forming 
H$_2$--bright hotspots.  Models investigating conditions in shocked gas with low metallicity are being investigated with this in mind (Guillard et al. 2016), and may help to motivate future far-IR attempts to detect rest-frame H$_2$ emission at redshifts at or beyond the era of re-ionization.   

\acknowledgements
This paper is dedicated to the work of James Houck of Cornell University  (1940-2015), who was the PI of the {\it Spitzer IRS} instrument.  Without his vision and dedication to building a superb instrument, the work described in this paper would never been possible. The authors wish to thank an anonymous referee for suggestions which helped to improve the paper. This work is based, in part, on observations (and archival observations) made with the {\it Spitzer}  Space Telescope, which is operated by the Jet Propulsion Laboratory, California Institute of Technology under a contract with NASA. The work 
is also based, in part, on observations made with {\it Herschel}, a European Space Agency Cornerstone Mission with significant participation by NASA. Partial support for the {\it Herschel}  work was provided by NASA through an award issued by JPL/Caltech.  UL acknowledges support by the research
projects AYA2011-24728 and AYA2014-53506-P  financed by the Spanish Ministerio de Econom\'ia
y Competividad and by FEDER (Fondo Europeo de Desarrollo Regional) and the Junta de Andaluc\'ia (Spain)
grants FQM108.

\newpage

\appendix

\counterwithin{figure}{section}
\counterwithin{table}{section}
\renewcommand{\thefigure}{A\arabic{figure}}
\renewcommand\thetable{A\arabic{table}}
\setcounter{figure}{0}
\setcounter{table}{0}

\section{Appendix}

{\bf Appendix 1 IRS Spectra of various extraction regions shown in Figure~\ref{spitzerfoot}}
\begin{figure*}[h]
\includegraphics*[width=1.0\textwidth]{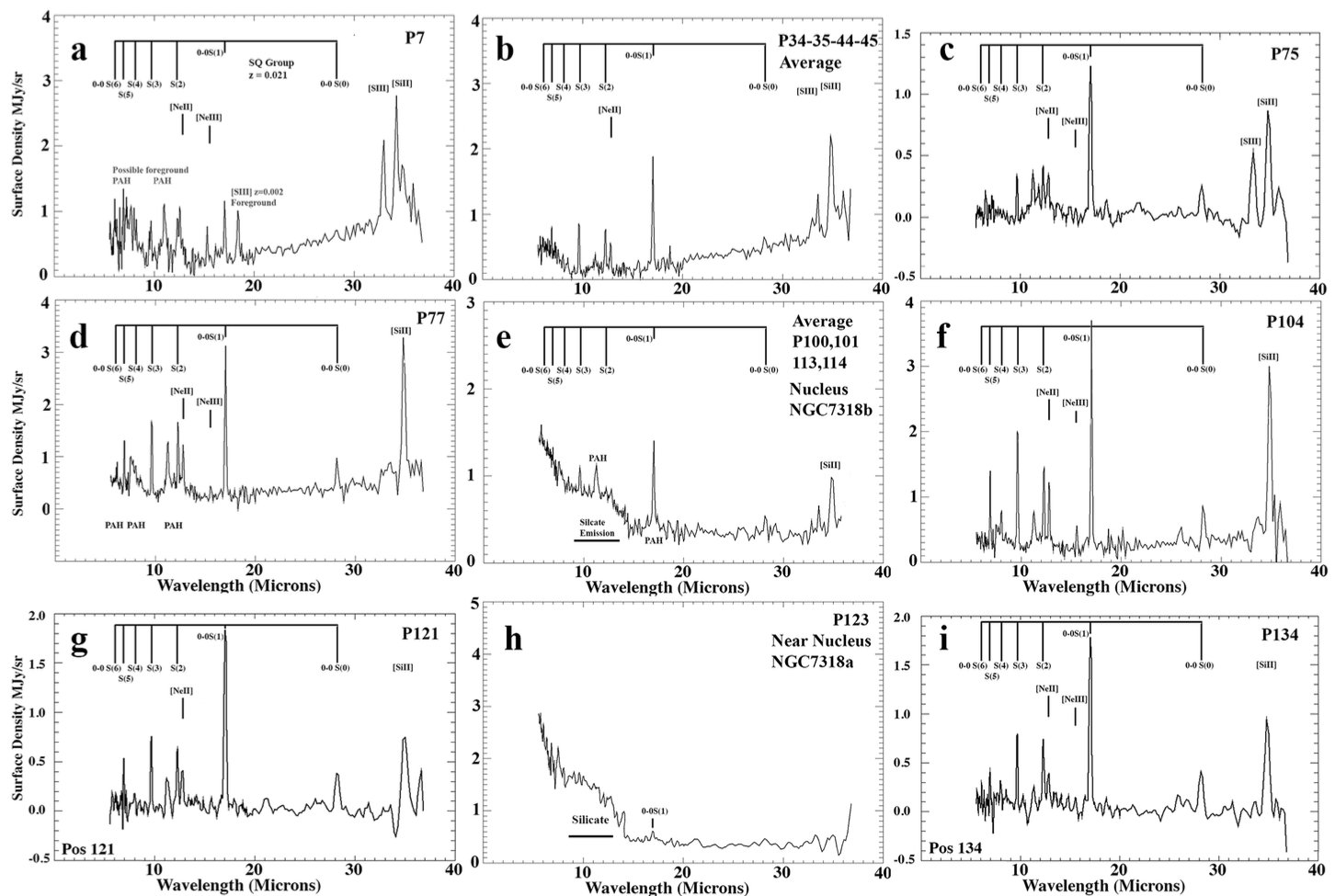}
\caption{Selected IRS spectra from spectral grid of Figure 1b with main detected lines identified. Spectra from: a) near bright HII region in foreground galaxy NGC 7320, position (P7). Note features of foreground and background group are mainly separated), b) average of four spectra centered on the high excitation region in the south (P34,35,44,45), c) behind main shock (P75), d) on main shock-south (P77), e) Nucleus of NGC 7318b (average P100,101,113,114), f) main shock-central (P104), g) AGN bridge region (P121), h) Near nucleus of NGC 7318a (P123), i) AGN bridge region (P134). }
\label{figa1}
\end{figure*}

\begin{figure*}[h]
\includegraphics*[width=1.0\textwidth]{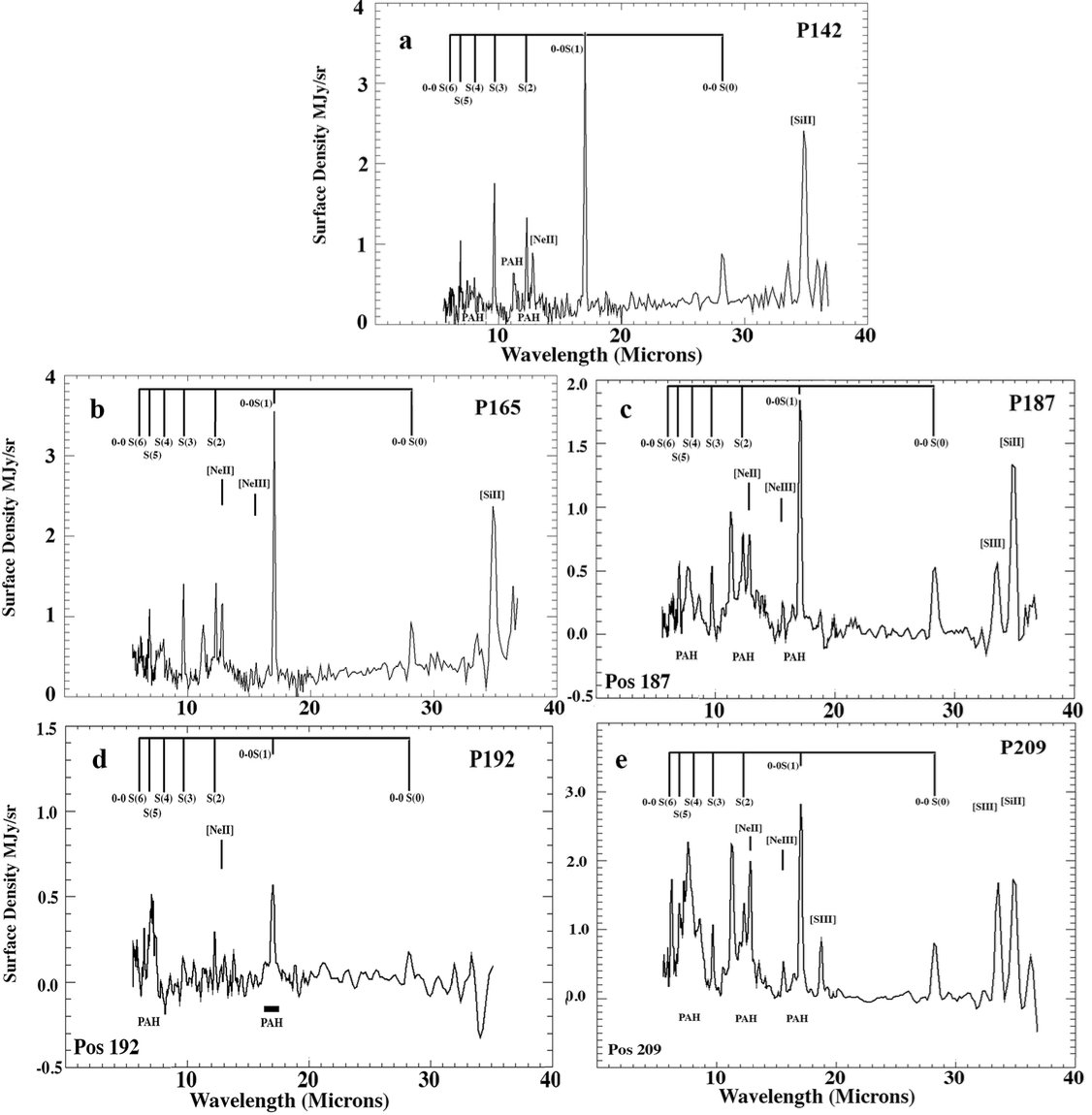}
\caption{b) Selected IRS spectra from spectral grid of Figure 1b with main detected lines identified. Spectra from: a) main shock-north (P142), b) main shock further north (P165), c) main shock and beginning of SQ-A complex -including rising PAH features indicating some heating from star formation (P187), d) In northern stellar loop from NGC 7318a (P192)-note very faint H$_2$ lines, e) SQ-A region (P209) and stronger PAHs.  }
\label{figa2}
\end{figure*}
\newpage

{\bf Appendix 2: Extrapolating the Mass of gas to low gas temperatures using the power--law method}

The power--law analysis described earlier can also make a prediction about the total molecular mass below 100K (the practical limit for exciting the lowest 0-0S(0) H$_2$ rotational line), since we can extrapolate the mass of gas below which the rotational H$_2$ emission is excited \citep[see][for more details]{togi16}.  This assumes that the same power--law derived for warmer gas can extrapolated to cooler temperatures.  Increasing the number of extremely cold H$_{2}$ molecules below a certain temperature, will cause a change in the total molecular gas mass, but will result in an insignificant change in the MIR H$_{2}$ rotational line flux. We define the sensitive temperature, T$_{s}$, below which H$_{2}$ molecules are too cold to cause substantial changes in the mid-IR rotational-line H$_2$ excitation diagram. 
In our analysis three different cases emerge in determining T$_{s}$ depending on the individual spectral region explored?

\begin{figure*}[h]
\centering
\includegraphics[width=1.0\textwidth]{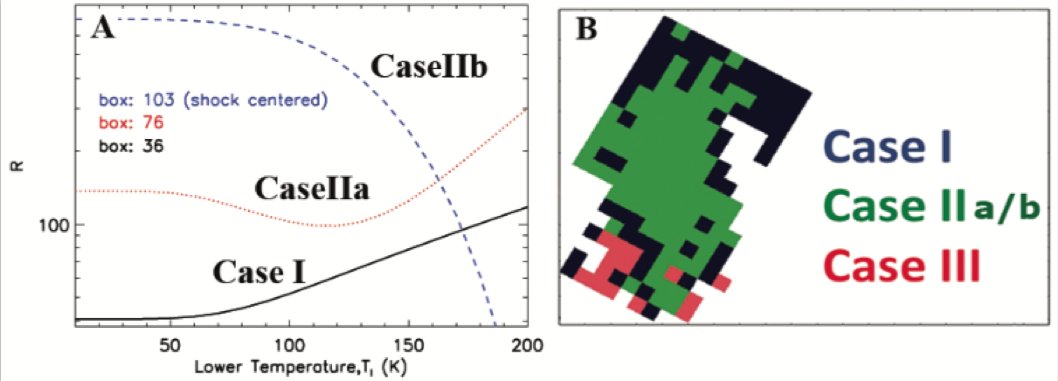}
\caption{ a) The difference between the model derived and the observed H$_{2}$ line ratios (R) as a function of lower temperature to determine T$_{\ell}$. The black solid, red dotted, and blue dashed curves represent different cases. For the black curve cases we adopt the value of T$_{\ell}$ = 50 K. The red and blue curve occur in warm regions, where the temperature of all H$_{2}$ molecules are high enough to be traced completely by the MIR rotational lines. In such cases we adopt the value of T$_{\ell}$ at the temperature where we have the minimum deviation. For instance for the red and blue curve we adopt T$_{\ell}$ = 120 and 198 K, respectively, given by our power--law model. b) The spatial distribution of the Case I, II and III extrapolation methods displayed as a schematic diagram covering the mapped area shown in Figure~\ref{spitzerfoot}b (see text). Case III are cases where the 0-0S(0) is not detected but the higher-J levels are, implying high temperatures.}
\label{figa3}
\end{figure*}

\begin{itemize}

\item{Case1:  We found that for some positions in the filament, as we decreased the lower temperature T$_{\ell}$, the difference between the observed and the model fit ratio (R) continuously decreases, and below a certain temperature (T$_{s}$) the deviation becomes insignificant, minor change in the model fit occurs. The parameter, $R$, is defined as,
\begin{equation}
R = \sum_{i=1}^m \left(\frac{f_{i,mod} - f_{i,obs}}{\sigma_{f_{i,obs}}}\right)^2,
\end{equation}
where $f = ln\left(\frac{N_{u}/g_{u}}{(N_{u}/g_{u})_{S(1)}}\right)$, f$_{i,mod}$ and f$_{i,obs}$ are the modeled and observed flux ratios for the $i^{th}$ line with uncertainty $\sigma_{f_{i,obs}}$, and the summation is over all the independent line flux ratios. The black solid curve in Figure \ref{figa3}a, demonstrate this case. In such scenario we extrapolate our power--law model to 50 K to calculate the total molecular gas mass. The calculated molecular gas mass by extrapolating our model to 50 K, correlates with the molecular gas mass derived using the Galactic-CO conversion factor, $\alpha_{CO,Gal}$ (Togi \& Smith, in prep)}

\item{Case2a: As we decrease the lower temperature, T$_{\ell}$, the value of $R$ decrease but below a certain temperature it starts increasing and at low temperatures show no significant change. The red dotted curve in the Figure \ref{figa3}a, demonstrate this case in the  region- 76. The value of $R$ decrease till the lower temperature of about 120 K but increase in the range 50--120 K and remains almost constant below 50 K. In such case we adopt the corresponding temperature for T$_{\ell}$, here 120 K, where the deviation is found to be minimum. The presence of excess warm molecular gas due to shocks could be the reason for high T$_{\ell}$ and the complete molecular gas in these regions are traced by the MIR H$_{2}$ rotational lines.

Case2b: The blue-dashed curve is a similar case but the minimum deviation occurs at very high T$_{\ell}$. This was observed in the Region 103 (see Figure~\ref{spitzerfoot}b), the central shocked region of the Quintet, where the model yield the best fit value of T$_{\ell}$ = 198 K. The molecular gas in this central shocked region of the Quintet is heated to a temperature of $\sim$ 200 K. }

\item{Case3: In some regions the S(0) line is undetected  but with detections of high J lines of H$_{2}$. The S(1) and higher J lines of H$_{2}$, correspond to energy levels, J $\geq$ 3. High J levels requires high temperatures for excitation, which result in high T$_{s}$. In such cases we extrapolate our power--law model to 80 K to calculate the total molecular gas mass. Several Ultra Luminous InfraRed Galaxies (ULIRGs), which are local mergers, with warm dust color temperatures have non detections of S(0) but good S/N (signal-to-noise) detection for high-J rotational lines \citep{Higdon06, Stierwalt14}. In these galaxies we need to extrapolate our power--law model till 80 K to recover the total molecular gas mass to be consistent with their dynamical mass estimates \citep{togi16}. }
\end{itemize}

We show in Figure~\ref{figa3}b a schematic map of the distribution of the various cases over the mapped area. The Case 1 (38$\%$) squares are dark and occupy mainly the cooler northern part of the mapped region and the median value of warm mass fraction is only 4$\%$.  Cases 2 (54$\%$ green squares) and Case 3 (8$\%$ red) occupy the regions where the stronger shocks dominate and have median warm mass fractions of 39 and 27$\%$ respectively.

Although we detect H$_2$ in almost every position observed, there are areas near the western and southern borders of the maps where no warm H$_2$ is detected, even though it lies within the mapping region. We note that there is potentially another possible scenario, where no rotational H$_{2}$ lines are detected at all because the gas has too low a temperature to be detected in the mid-IR.  Although this is theoretically possible,  we assume instead the simpler explanation, that there is no molecular gas at these undetected positions.  This is probably justifiable, given that both single-dish and interferometric observations confine the CO-emitting gas to similar regions to the warm H$_2$ distribution. Furthermore, the C+ emission (Figure~\ref{Carbon2}), which could potential map CO-dark molecular gas \citep{wol10}, also seems to follow the same spatial distribution as the warm H$_2$ emission, and so large quantities of intergalactic "dark" molecules are not strongly favored by the current observations.  

\newpage

\begin{longtable}{lcccccccccc}
\tabletypesize{\scriptsize}
\tablecaption{Extracted Line Properties for Each Region shown in Figure~\ref{spitzerfoot}b. Line fluxes in units of $\times$ 10$^{-18}$W m$^{-2}$}
\tablewidth{0pt}
\label{taba1}
\tablehead{
\colhead{Region} & \colhead{RA (J2000)} & \colhead{Dec(J2000)} & \colhead{0-0S(0) (unc)} &  \colhead{0-0S(1) (unc)} & \colhead{0-0S(2) (unc)} & \colhead{0-0S(3) (unc)} &  \colhead{0-0S(4) (unc)} & \colhead{0-0S(5) (unc)} & \colhead{0-0S(6) (unc)} & \colhead{Detected } \\
\colhead{(Fig.~\ref{spitzerfoot}b)\tablenotemark{a}} & \colhead{(deg)} & \colhead{(deg)} & ($\lambda$28.2$\mu$m)\tablenotemark{b}  &($\lambda$17.0$\mu$m)\tablenotemark{b}  & ($\lambda$12.3$\mu$m)\tablenotemark{b} & ($\lambda$9.7$\mu$m)\tablenotemark{b} & ($\lambda$8.0$\mu$m)\tablenotemark{b}  & ($\lambda$6.9$\mu$m)\tablenotemark{b} & ($\lambda$6.1$\mu$m)\tablenotemark{b} & \colhead{Lines\tablenotemark{c}}\\
}
\startdata
1 & 338.99647 & 33.95042 & $<$0.13\tablenotemark{d} & $<$0.17 & $<$0.25 & $<$0.38 & $<$0.56 & $<$0.39 & $<$0.39 & no--lines \\
2 & 338.99808 & 33.95118 & $<$0.13 & 0.25 (0.06) & 0.40 (0.07) & $<$0.38 & $<$0.56 & 0.79 (0.04) & $<$0.39 & 1 \\
3 & 338.99970 & 33.95194 & $<$0.13 & 0.47 (0.06) & 0.71 (0.10) & $<$0.38 & $<$0.56 & $<$0.39 & $<$0.39 & 12 \\
4 & 339.00132 & 33.95270 & 0.11 (0.04) & 1.32 (0.06) & 1.01 (0.08) & 3.05 (0.13)  & $<$0.56 & $<$0.39 & $<$0.39 & 123 \\
5 & 339.00293 & 33.95346 & $<$0.13 & 1.21 (0.06) & 0.66 (0.16) & 2.13 (0.13) & $<$0.56 & $<$0.39 & $<$0.39 & 123 \\
6 & 339.00455 & 33.95422 & $<$0.13 & 0.91 (0.06) & $<$0.25 & $<$0.38 & $<$0.56 & 0.29 (0.04) & $<$0.39 & 1 \\
7 & 339.00617 & 33.95498 & $<$0.13 & 1.06 (0.06) & 0.85 (0.40) & 1.77 (0.13) & $<$0.56 & 2.18 (0.30) & $<$0.39 & 135 \\
8 & 339.00778 & 33.95574 & $<$0.13 & 1.11 (0.05) & $<$0.25 & 2.14 (0.13) & $<$0.56 & 0.86 (0.05) & $<$0.39 & 135 \\
\enddata
\tablenotetext{a}{Extracted over regions defined in Figure~\ref{spitzerfoot}b and of area 5.1 x 5.1 arcsecs$^2$. }
\tablenotetext{b}{1-$\sigma$ uncertainties derived from PAHFIT.}
\tablenotetext{c}{Lines detected 0=0-0S(0),1=0-0S(1) etc. Square brackets denote "marginal detection" used to help constrain 2-Temperature fits in a minority of cases. Those fluxes are also denoted by square brackets.}
\tablenotetext{d}{3-$\sigma$ upper limits.}
\end{longtable}

\newpage
\renewcommand\thetable{A\arabic{table}}

\newpage

\begin{longtable}{lccccccccc}
\tabletypesize{\scriptsize}
\tablecaption{Results from One or Two-temperature fitting\label{2temp} of the Warm Molecular Hydrogen}
\tablewidth{0pt}
\tablehead{
\colhead {Region\tablenotemark{a}}  & \colhead{T$_1$ (unc)\tablenotemark{b}}  & \colhead {OPR$_1$}  & \colhead {N(H$_2$)$_1$}               & \colhead{$\Sigma$(H$_2$)$_1$ (unc)\tablenotemark{b}} & \colhead{T$_2$ (unc)\tablenotemark{b} }     & \colhead {OPR$_2$}    & \colhead {N(H$_2$)$_2$}     & \colhead{$\Sigma$(H$_2$)$_2$ (unc)\tablenotemark{b}} & \colhead{Lines} \\
\colhead{} 					& \colhead{(K)} 	       &  				    & \colhead{(10$^{19}$ cm$^{-2}$)}    & \colhead{$\times$10$^6$ M$_{\odot}$ kpc$^{-2}$} &  \colhead{(K)} 	       &  				    & \colhead{(10$^{17}$ cm$^{-2}$)}    & \colhead{$\times$10$^{4}$M$_{\odot}$ kpc$^{-2}$} & (0-0S(...)) \\
}
\startdata
 
    3 &   154-805\tablenotemark{c} &   2.5-3.0\tablenotemark{c} &  2.5-0.05\tablenotemark{c}  &    0.4-0.08\tablenotemark{c} &  --- &     --- &    --- &  --- &    12  \\
    4  &  223-425\tablenotemark{c} &   2.9-3.0\tablenotemark{c} &   1.13-0.24\tablenotemark{c} &   0.18-0.04\tablenotemark{c} &  --- &     --- &   --- &  --- &   123  \\
    5  &  205-361\tablenotemark{c} &   2.9-3.0\tablenotemark{c} &   1.44-0.28\tablenotemark{c} &   0.23-0.04\tablenotemark{c} &  --- &     --- &    --- &  --- &   123  \\
    8  &  199-439\tablenotemark{c} &   2.8-3.0\tablenotemark{c} &   1.51-0.18\tablenotemark{c} &  0.24-0.03\tablenotemark{c} &  --- &     --- &   --- &  --- &   135  \\
    9  &  187-426\tablenotemark{c} &   2.79-3.0\tablenotemark{c} &   1.69-0.2\tablenotemark{c} &   0.27-0.03\tablenotemark{c} &  --- &     --- &   --- &  --- &   123  \\
   10  &  154-398\tablenotemark{c} &  2.54-2.99\tablenotemark{c} &   4.3-0.2\tablenotemark{c} &   0.69-0.3\tablenotemark{c} &  --- &     --- &    --- &  --- &   135  \\
   12  &  184-372\tablenotemark{c} &  2.77-3.0\tablenotemark{c} &  1.74-0.2\tablenotemark{c}  &    0.28-0.03\tablenotemark{c} &  --- &     --- &   --- &  --- &   123  \\
13 & 227 (34) & 2.94 &  1.67 &  0.27 (0.16) &  732 (73)& 3.0&  2.50 & 0.39 (0.08) & 0123[5]\tablenotemark{d} \\
\enddata
\tablenotetext{a}{Regions defined from Figure~\ref{spitzerfoot}b.}
\tablenotetext{b}{1-$\sigma$ uncertainties shown in brackets}
\tablenotetext{c}{Not enough points to fit two temperatures. Range based on single temperature fits to lowest 3 transitions with and without the 3-sigma S(0) upper limit.} 
\tablenotetext{d}{Marginal detections of 0-0S(5) line provides additional information to allow 2-temperature fit.} 
\label{taba2}
\end{longtable}

\newpage

\begin{longtable*}{cccccc|cccccc}
\centering
\tabletypesize{\scriptsize}
\tablecaption{Power law model results for Quintet}
\tablewidth{0pt}
\tablecolumns{12} 
\tablehead{
\colhead{Region}  & \colhead{n($\delta$n)} & \colhead{$T_{\ell}$} & M($>$ 120 K) & \colhead{$\frac{M(T>120 K)}{M_{total}}$} & \colhead{Class} & \colhead{Region}  & \colhead{n($\delta$n)} & \colhead{$T_{\ell}$} & M($>$ 120 K) & \colhead{$\frac{M(T>120 K)}{M_{total}}$} & \colhead{Class}\\
\colhead{}  & \colhead{}  & \colhead{(K)} & \colhead{($10^{6}M_{\odot}$)} & \colhead{($\%$)} &\colhead{Type} & \colhead{}  & \colhead{}  & \colhead{(K)} & \colhead{($10^{6}M_{\odot}$)} & \colhead{($\%$)} &\colhead{Type}
}

4   & 3.44 (0.07)    & 50      & 1.27  &   11.8 & I   & 109 & 4.78 (0.04)    & 100     & 6.58    & 50.2  & II\\
5   & 3.84 (3.84)    & 80      & 1.49  &   31.7 & III & 113 & 4.19 (0.51)    & 50      & 1.20    &  6.1  & I\\
7   & 3.78 (3.78)    & 80      & 1.26  &   32.4 & III & 114 & 4.52 (0.13)    & 50      & 2.12    &  4.6 & I\\
8   & 4.19 (4.19)    & 80      & 1.67  &   27.4 & III & 115 & 4.57 (0.08)    & 100     & 4.47    & 52.2 & II\\
9   & 3.65 (3.65)    & 80      & 1.01  &   34.1 & III & 116 & 4.25 (0.03)    & 100     & 6.68    & 55.3 & II\\
10  & 4.37 (0.12)    & 50      & 1.33  &   5.22 & I   & 117 & 4.20 (0.02)    & 110     & 6.84    & 75.7 & II\\
12  & 4.24 (4.24)    & 80      & 1.35  &   26.9 & III & 118 & 4.40 (0.03)    & 100     & 6.28    & 53.8 & II\\
13  & 4.62 (0.12)    & 110     & 4.23  &    73  & II  & 119 & 4.59 (0.04)    & 90      & 6.00    & 35.6 & II\\
\end{longtable*}


\begin{longtable}{llllllll}
\label{taba4}
\tabletypesize{\scriptsize}
\tablecaption{H$_2$ fluxes fitted with shock models with pre-shock densities  $n_{\rm H}$=10$^4$~cm$^{-3}$ and 10$^3$~cm$^{-3}$ for the slow and fast shocks respectively.}
\tablewidth{0pt}
\tablehead{
\colhead{Region} & \colhead{V$_{\rm s}$\tablenotemark{a}} & \colhead{Mass flow (120K)\tablenotemark{b} }& \colhead{t$_{\rm cool}$(120K)\tablenotemark{c}} &  \colhead{$M(120K)($H$_2)$\tablenotemark{d}} & \colhead{Mass flow (50K)\tablenotemark{b} }& \colhead{t$_{\rm cool}$(50K)\tablenotemark{c}} &  \colhead{$M(50K)($H$_2)$\tablenotemark{d}} \\
\colhead{} & \colhead{[km s$^{-1}$]} & \colhead{[10$^2$ M$_{\odot}$ yr$^{-1}$]} & \colhead{[10$^2$ yrs]} & \colhead{[10$^5$ M$_{\odot}$]} & \colhead{[10$^2$M$_{\odot}$ yr$^{-1}$]} & \colhead{[10$^2$ yrs]} & \colhead{[10$^5$ M$_{\odot}$]}\\
}
\startdata

 \multirow{2}*{   1 }   &      6.0 &      3.25~(0.86)   &      2.47 &      0.80 (0.21) &      3.23 (0.86)   &      5.08 &      1.64 (0.44)  \\ 
                                 &    33.5 &      3.51~(0.77)   &     10.86 &      3.81 (0.84) &      2.17 (0.48)   &     33.27 &      7.20 (1.59)  \\ 
\hline
 \multirow{2}*{   2 }   &      5.5 &      6.71~(1.23)   &      2.30 &      1.55 (0.28) &      6.80 (1.22)   &      5.42 &      3.68 (0.66)  \\ 
                                 &    32.5 &      4.75~(0.26)   &      6.36 &      3.02 (0.17) &      2.93 (0.16)   &     32.25 &      9.44 (0.52)  \\ 
\hline
 \multirow{2}*{   3 }   &      5.5 &     12.42 (1.32)   &      2.30 &      2.86 (0.31) &     12.43 (1.32)   &      5.42 &      6.74 (0.71)  \\ 
                                 &    32.5 &      2.92 (0.65)  &      6.36 &      1.85 (0.42) &      1.76 (0.40)   &     32.25 &      5.69 (1.29)  \\ 
\hline
 \multirow{2}*{   4 }   &      7.0 &     15.25 (0.49)   &      2.63 &      4.02 (0.13) &     15.22 (0.49)   &      4.86 &      7.40 (0.24)  \\ 
                                 &    33.5 &      2.97   (0.81)   &     10.86 &      3.22 (0.88) &      1.84 (0.50)   &     33.27 &      6.14 (1.67)  \\
\enddata
\end{longtable}


\begin{thebibliography}{}


\bibitem[Alatalo et al.(2013)]{ala13} Alatalo, K., Davis,  T.~A., Bureau, M., et al.\ 2013, \mnras, 432, 1796 

\bibitem[Alatalo et al.(2014)]{ala14} Alatalo, K., Appleton, P.~N., Lisenfeld, U., et al.\ 2014, \apj, 795, 159  

\bibitem[Allen  \& Hartsuiker(1972)]{all72} Allen, R.~J., \& Hartsuiker, J.~W.\ 1972, \nat, 239, 324

\bibitem[Appleton et al.(2006)]{app06} Appleton, P.~N., Xu, K.~C., Reach, W., et al.\ 2006, \apjl, 639, L51
\bibitem[Appleton et al.(2009)]{app09} Appleton, P., Armus, L., Blain, A., et al.\ 2009, astro2010: The Astronomy and Astrophysics Decadal Survey, 2010
\bibitem[Appleton et al.(2013)]{app13} Appleton, P.~N.,  Guillard, P., Boulanger, F., et al.\ 2013, \apj, 777, 66 

\bibitem[Bahcall et al.(1984)]{bah84} Bahcall, N.~A., Harris,  D.~E., \& Rood, H.~J.\ 1984, \apjl, 284, L29 

\bibitem[Bitsakis et  al.(2010)]{bit10} Bitsakis, T., Charmandaris, V., Le Floc'h, E., et al.\ 2010, \aap, 517, A75 

\bibitem[Bitsakis et al.(2011)]{bit11} Bitsakis, T., Charmandaris, V., da Cunha, E., et al.\ 2011, \aap, 533, A142

\bibitem[Bitsakis et  al.(2014)]{bit14} Bitsakis, T., Charmandaris, V., Appleton, P.~N., et al.\ 2014, \aap, 565, A25


\bibitem[Browning et al.(2003)]{Browning03} Browning, M.~K., Tumlinson, J., \& Shull, J.~M.\ 2003, \apj, 582, 810 

\bibitem[Burton(1987)]{Burton87} Burton, M.~G.\ 1987, Ph.D.~Thesis

\bibitem[Cluver et al.(2010)]{clu10} Cluver, M.~E., Appleton,  P.~N., Boulanger, F., et al.\ 2010, \apj, 710, 248 (Paper 1)

\bibitem[Cluver et al.(2013)]{clu13} Cluver, M.~E., Appleton,  P.~N., Ogle, P., et al.\ 2013, \apj, 765, 93

\bibitem[Donahue et al.(2011)]{don11} Donahue, M., de Messi{\`e}res, G.~E., O'Connell, R.~W., et al.\ 2011, \apj, 732, 40

\bibitem[Egami et al.(2006)]{ega06} Egami, E., Rieke, G.~H., Fadda, D., \& Hines, D.~C.\ 2006, \apjl, 652, L21 

\bibitem[Fedotov et al.(2011)]{fed11} Fedotov, K., Gallagher,  S.~C., Konstantopoulos, I.~S., et al.\ 2011, \aj, 142, 42

\bibitem[Fiolet et al.(2010)]{fio10}Fiolet, N., Omont, A., Lagache, G., et al. 2010, \aap, 524, 33


\bibitem[Gallagher et al.(2001)]{gal01} Gallagher, S.~C.,  Charlton, J.~C., Hunsberger, S.~D., Zaritsky, D.,  \& Whitmore, B.~C.\ 2001, \aj, 122, 163

\bibitem[Geng et al.(2012)]{gen12} Geng, A., Beck, A.~M.,  Dolag, K., et al.\ 2012, \mnras, 426, 3160 

\bibitem[Gallagher et al.(2001)]{gal01} Gallagher, S.~C., Charlton, J.~C., Hunsberger, S.~D., Zaritsky, D., \& Whitmore, B.~C.\ 2001, \aj, 122, 163


\bibitem[Gao  \& Xu(2000)]{gao00} Gao, Y., \& Xu, C.\ 2000, \apjl, 542, L83

\bibitem[Guillard et al.(2009)]{gui09} Guillard, P., Boulanger, F., Pineau Des For{\^e}ts, G., \& Appleton, P.~N.\ 2009, \aap, 502, 515

\bibitem[Guillard et  al.(2010)]{gui10} Guillard, P., Boulanger, F., Cluver, M.~E., et al.\ 2010, \aap, 518, A59

\bibitem[Guillard et al.(2012)]{gui12} Guillard, P., Boulanger, F., Pineau des For{\^e}ts, G., et al.\ 2012, \apj, 749, 158 
\bibitem[Guillard et al.(2012b)]{gui12b} Guillard, P., Ogle, P.~M., Emonts, B.~H.~C., et al.\ 2012, \apj, 747, 95
\bibitem[Guillard et al.(2015)]{gui15} Guillard, P., Boulanger, F., Lehnert, M.~D., Appleton, P.~N., \& Pineau des For{\^e}ts, G.\ 2015, SF2A-2015: Proceedings of the Annual meeting of the French Society of Astronomy and Astrophysics, 81 
\bibitem[Guillard et al (2016)]{gui16} Guillard, P. Appleton, P. N., Boulanger, F. et al.\ 2016, \aap, submitted. 
\bibitem[Hatziminaoglou et al.(2015)]{hat15} Hatziminaoglou, E., Hern{\'a}n-Caballero, A., Feltre, A., \& Pi{\~n}ol Ferrer, N.\ 2015, \apj, 803, 110
\bibitem[Higdon et al.(2006)]{Higdon06} Higdon, S.~J.~U., Armus, L., Higdon, J.~L., Soifer, B.~T., \& Spoon, H.~W.~W.\ 2006, \apj, 648, 323 

\bibitem[Hollenbach \& McKee(1979)]{Hollenbach79} Hollenbach, D., \& McKee, C.~F.\ 1979, \apjs, 41, 555 

\bibitem[Hwang et al.(2012)]{hwa12} Hwang, J.-S., Struck, C.,  Renaud, F., \& Appleton, P.~N.\ 2012, \mnras, 419, 1780

\bibitem[Ingalls et al.(2011)]{Ingalls11} Ingalls, J.~G., Bania, T.~M., Boulanger, F., et al.\ 2011, \apj, 743, 174 

\bibitem[Iglesias-P{\'a}ramo et  al.(2012)]{igl12} Iglesias-P{\'a}ramo, J., L{\'o}pez-Mart{\'{\i}}n, L., V{\'{\i}}lchez, J.~M., Petropoulou, V., \& Sulentic, J.~W.\ 2012, \aap, 539, A127 

\bibitem[Konstantopoulos et al.(2014)]{kon14}  Konstantopoulos, I.~S., Appleton, P.~N., Guillard, P., et al.\ 2014, \apj, 
784, 1 
\bibitem[Lanz et al.(2015)]{lan15} Lanz, L., Ogle, P.~M., Evans, D., et al.\ 2015, \apj, 801, 17 
\bibitem[Le Bourlot et al.(2002)]{Leb02} Le Bourlot, J., Pineau des For{\^e}ts, G., Flower, D.~R., \& Cabrit, S.\ 2002, \mnras, 332, 985 
\bibitem[Lesaffre et  al.(2013)]{les13} Lesaffre, P., Pineau des For{\^e}ts, G., Godard, B., et al.\ 2013, \aap, 550, A106 
\bibitem[Li et al.(2008)]{li08} Li, M.~P., Shi, Q.~J., \& Li, A.\ 2008, \mnras, 391, L49   \bibitem[Lisenfeld et  al.(2002)]{lis02} Lisenfeld, U., Braine, J., Duc, P.-A., et al.\ 2002, \aap, 394, 823
\bibitem[Sault et al.(1995)]{miriad} Sault, R.~J., Teuben,  P.~J.,  \& Wright, M.~C.~H.\ 1995, Astronomical Data Analysis Software and Systems IV, 77, 433 

\bibitem[Moles et al.(1997)]{mol97} Moles, M., Sulentic,  J.~W., \& M{\'a}rquez, I.\ 1997, \apjl, 485, L69 


\bibitem[Moles et  al.(1998)]{mol98} Moles, M., Marquez, I., \& Sulentic, J.~W.\ 1998, \aap, 334, 473

\bibitem[Natale et al.(2010)]{nat10} Natale, G., Tuffs,  R.~J., Xu, C.~K., et al.\ 2010, \apj, 725, 955
\bibitem[Nesvadba et al.(2010)]{nes10} Nesvadba, N.~P.~H., Boulanger, F., Salom{\'e}, P., et al.\ 2010, \aap, 521, A65 
\bibitem[Nesvadba et al.(2011)]{nes11} Nesvadba, N.~P.~H., Boulanger, F., Lehnert, M.~D., Guillard, P., \& Salome, P.\ 2011, \aap, 536, L5 
\bibitem[Neufeld \& Yuan(2008)]{Neufeld08} Neufeld, D.~A., \& Yuan, Y.\ 2008, \apj, 678, 974

\bibitem[Nikiel-Wroczy{\'n}ski et al.(2013)]{nik13}  Nikiel-Wroczy{\'n}ski, B., Soida, M., Urbanik, M., Beck, R.,  \& Bomans, D.~J.\ 2013, \mnras, 435, 149

\bibitem[O'Sullivan et al.(2009)]{osu09} O'Sullivan, E.,  Giacintucci, S., Vrtilek, J.~M., Raychaudhury, S.,  \& David, L.~P.\ 2009, \apj, 701, 1560
\bibitem[Ogle et al.(2007)]{ogl07} Ogle, P., Antonucci, R., Appleton, P.~N., \& Whysong, D.\ 2007, \apj, 668, 699 
\bibitem[Ogle et al.(2010)]{ogl10} Ogle, P., Boulanger, F., Guillard, P., Evans, D.A., Antonucci, R., Appleton, P.N, Nesvadba, N. \& Leipski, C. 2010, \apj, 724, 1193
\bibitem[Ogle et al.(2012)]{ogl12} Ogle, P., Davies, J.~E., Appleton, P.~N., et al.\ 2012, \apj, 751, 13 
\bibitem[Ogle, Lanz \& Appleton (2014)]{ogl14} Ogle, P.~M., Lanz, L.,  \& Appleton, P.~N.\ 2014, \apj (In Press) and  arXiv:1405.2040 

\bibitem[Peterson et al.(2012)]{pet12} Peterson, B.~W.,  Appleton, P.~N., Helou, G., et al.\ 2012, \apj, 751, 11 

\bibitem[Petitpas  \& Taylor(2005)]{pet05} Petitpas, G.~R., \& Taylor, C.~L.\ 2005, \apj, 633, 138


\bibitem[Pilbratt et  al.(2010)]{pil10} Pilbratt, G.~L., Riedinger, J.~R., Passvogel, T., et al.\ 2010, \aap, 518, L1 

\bibitem[Poglitsch et al.(2010)]{pog10} Poglitsch, A., Waelkens, C., Geis, N., et al.\ 2010, \aap, 518, L2 

\bibitem[Rachford et al.(2002)]{Rachford02} Rachford, B.~L., Snow, T.~P., Tumlinson, J., et al.\ 2002, \apj, 577, 221 

\bibitem[Renaud et al.(2010)]{ren10} Renaud, F., Appleton,  P.~N., \& Xu, C.~K.\ 2010, \apj, 724, 80

\bibitem[Rigopoulou et al.(2002)]{rig02} Rigopoulou, D., Kunze, D., Lutz, D., Genzel, R., \& Moorwood, A.~F.~M.\ 2002, \aap, 389, 374 

\bibitem[Rodr{\'{\i}}guez-Baras et al.(2014)]{rod14}  Rodr{\'{\i}}guez-Baras, M., Rosales-Ortega, F.~F., D{\'{\i}}az, A.~I.,  S{\'a}nchez, S.~F., \& Pasquali, A.\ 2014, \mnras, 442, 495

\bibitem[Roussel et al.(2007)]{rou07} Roussel, H., Helou, G.,  Hollenbach, D.~J., et al.\ 2007, \apj, 669, 959 

\bibitem[Roussel(2013)]{rou13} Roussel, H.\ 2013, \pasp, 125,  1126

\bibitem[Savage et al.(1977)]{Savage77} Savage, B.~D., Bohlin, R.~C., Drake, J.~F., \& Budich, W.\ 1977, \apj, 216, 291 

\bibitem[Smith  \& Struck(2001)]{smi01} Smith, B.~J., \& Struck, C.\ 2001, \aj, 121, 710

\bibitem[Smith et al.(2007a)]{smi07a} Smith, J.~D.~T., Armus,  L., Dale, D.~A., et al.\ 2007a, \pasp, 119, 1133 

\bibitem[Smith et al.(2007b)]{smi07b} Smith, J.~D.~T., Draine,  B.~T., Dale, D.~A., et al.\ 2007b, \apj, 656, 770 

\bibitem[Snow \& McCall(2006)]{Snow06} Snow, T.~P., \& McCall, B.~J.\ 2006, \araa, 44, 367

\bibitem[Spitzer \& Zweibel(1974)]{Spitzerzweibel74} Spitzer, L., Jr., \& Zweibel, E.~G.\ 1974, \apjl, 191, L127 

\bibitem[Spitzer et al.(1974)]{Spitzer74} Spitzer, L., Jr., Cochran, W.~D., \& Hirshfeld, A.\ 1974, \apjs, 28, 373 

\bibitem[Spitzer \& Cochran(1973)]{Spitzercoch73} Spitzer, L., Jr., \& Cochran, W.~D.\ 1973, \apjl, 186, L23 

\bibitem[Spitzer et al.(1973)]{Spitzer73} Spitzer, L., Drake, J.~F., Jenkins, E.~B., et al.\ 1973, \apjl, 181, L116 

\bibitem[Strateva et al.(2001)]{str01} Strateva, I.,  Ivezi{\'c}, {\v Z}., Knapp, G.~R., et al.\ 2001, \aj, 122, 1861 

\bibitem[Stierwalt et al.(2014)]{Stierwalt14} Stierwalt, S., Armus, L., Charmandaris, V., et al.\ 2014, \apj, 790, 124 

\bibitem[Struck (1997)]{str97}Struck, C. 1997, ApJS, 113, 269

\bibitem[Sulentic et  al.(1995)]{sul95} Sulentic, J.~W., Pietsch, W., \& Arp, H.\ 1995, \aap, 298, 420

\bibitem[Sulentic et al.(2001)]{sul01} Sulentic, J.~W.,  Rosado, M., Dultzin-Hacyan, D., et al.\ 2001, \aj, 122, 2993

\bibitem[Suzuki et al.(2011)]{suz11} Suzuki, T., Kaneda, H.,  Onaka, T., \& Kitayama, T.\ 2011, \apjl, 731, L12

\bibitem[Trinchieri et  al.(2003)]{tri03} Trinchieri, G., Sulentic, J., Breitschwerdt, D., \& Pietsch, W.\ 2003, \aap, 401, 173

\bibitem[Togi \& Smith(2016)]{togi16} Togi, A. \& Smith, J. D.\ 2016, \apj, (Submitted). 

\bibitem[Valentijn \& van der Werf(1999)]{Valentijn99} Valentijn, E.~A., \& van der Werf, P.~P.\ 1999, \apjl, 522, L29

\bibitem[van der Hulst  \& Rots(1981)]{van81} van der Hulst, J.~M., \& Rots, A.~H.\ 1981, \aj, 86, 1775

\bibitem[Williams et al.(2002)]{wil02} Williams, B.~A., Yun,  M.~S., \& Verdes-Montenegro, L.\ 2002, \aj, 123, 2417

\bibitem[Wolfire et al.(2010)]{wol10} Wolfire, M.~G., Hollenbach, D., \& McKee, C.~F.\ 2010, \apj, 716, 1191 

\bibitem[Xu et al.(1999)]{xu99} Xu, C., Sulentic, J.~W.,  \& Tuffs, R.\ 1999, \apj, 512, 178 

\bibitem[Xu et al.(2003)]{xu03} Xu, C.~K., Lu, N., Condon,  J.~J., Dopita, M., \& Tuffs, R.~J.\ 2003, \apj, 595, 665

\bibitem[Xu et al.(2005)]{xu05} Xu, C.~K.,  Iglesias-P{\'a}ramo, J., Burgarella, D., et al.\ 2005, \apjl, 619, L95

\bibitem[Yun et al.(1997)]{yun97} Yun, M.~S.,  Verdes-Montenegro, L., del Olmo, A., \& Perea, J.\ 1997, \apjl, 475, L21


\end{thebibliography}
\end{document}